\documentclass[aps,pra,twocolumn,superscriptaddress]{revtex4}% review and submission% review and submission
\pdfoutput=1
\usepackage[pdftex]{graphicx}
\usepackage{dcolumn}   % needed for some tables
\usepackage{bm}        % for math
\usepackage{amssymb}   % for math
\usepackage{verbatim}
\usepackage[T1]{fontenc}

\usepackage{graphicx}
\usepackage{epstopdf}
\usepackage{subfigure}
\usepackage[urlcolor=blue,hyperindex,colorlinks,bookmarks=true,linkcolor=black,citecolor=black]{hyperref}
\usepackage[normalem]{ulem}
\usepackage[abs]{overpic}
\usepackage{color}
\usepackage{rotating}

\usepackage{amsmath}
\usepackage{amsfonts}
\usepackage{float}

\newcommand{\bra}[1]{\langle #1|}
\newcommand{\ket}[1]{|#1\rangle}

\def\lsim{\mathrel{\rlap{\lower4pt\hbox{\hskip1pt$\sim$}}
    \raise1pt\hbox{$<$}}}                % less than or approx. symbol
\def\gsim{\mathrel{\rlap{\lower4pt\hbox{\hskip1pt$\sim$}}
    \raise1pt\hbox{$>$}}}                % greater than or approx. symbol
\newcommand{\be}{\begin{equation}}
\newcommand{\ee}{\end{equation}}
\newcommand{\bea}{\begin{align}}
\newcommand{\eea}{\end{align}}

\newcommand{\abs}[1]{\lvert#1\rvert}

\newcommand{\yy}{\gamma}
\newcommand{\al}{\alpha}
\newcommand{\ww}{\omega}

\begin{document}

\title {High-fidelity qubit measurement with a microwave photon counter}

\author{Luke C.G. Govia}
\email[Electronic address: ]{lcggovia@lusi.uni-sb.de}
\affiliation{Theoretical Physics, Saarland University, Campus, 66123 Saarbr\"ucken, Germany}

\author{Emily J. Pritchett}
\affiliation{HRL Laboratories, LLC, Malibu, CA 90265, USA}

\author{Canran Xu}
\affiliation{Department of Physics, University of Wisconsin, Madison, WI 53706, USA}

\author{B.L.T. Plourde}
\affiliation{Department of Physics, Syracuse University, Syracuse, NY 13244-1130, USA}

\author{Maxim G. Vavilov}
\affiliation{Department of Physics, University of Wisconsin, Madison, WI 53706, USA}

\author{Frank K. Wilhelm}
\affiliation{Theoretical Physics, Saarland University, Campus, 66123 Saarbr\"ucken, Germany}

\author{R. McDermott}
\affiliation{Department of Physics, University of Wisconsin, Madison, WI 53706, USA}

\date{\today}

\begin{abstract}
High-fidelity, efficient quantum nondemolition readout of quantum bits is integral to the goal of quantum computation.  As superconducting circuits approach the requirements of scalable, universal fault tolerance, qubit readout must also meet the demand of simplicity to scale with growing system size.  Here we propose a fast, high-fidelity, scalable measurement scheme based on the state-selective ring-up of a cavity followed by photodetection with the recently introduced Josephson photomultiplier (JPM), a current-biased Josephson junction.   This scheme maps qubit state information to the binary digital output of the JPM, circumventing the need for room-temperature heterodyne detection and offering the possibility of a cryogenic interface to superconducting digital control circuitry.  Numerics show that measurement contrast in excess of 95\% is achievable in a measurement time of 140 ns. We discuss perspectives to scale this scheme to enable readout of multiple qubit channels with a single JPM.
\end{abstract}

\maketitle

\section*{{\normalsize{\textbf{I. Introduction}}}}

Over the past decade, circuit quantum electrodynamics (cQED) has emerged as a powerful paradigm for scalable quantum information processing in the solid state \cite{Blais04, Wallraff04, Wallraff05, Schuster05}. Here a superconducting qubit plays the role of an artificial atom, and a thin-film coplanar waveguide or bulk cavity resonator is used to realize a bosonic mode with strong coupling to the atom. Interaction between the qubit and the cavity is described by the Jaynes-Cummings Hamiltonian \cite{Jaynes63}. Strong interaction between the qubit and the cavity has been used to realize high-fidelity multi-qubit gates \cite{Majer07,DiCarlo09,DiCarlo10,Chow12}; moreover, the qubit has been used to prepare highly nonclassical states of the resonator \cite{Hofheinz08,Hofheinz09}. In the limit where the qubit is far detuned from the cavity resonance so that $\Delta \equiv \omega_{\rm C}-\omega_{\rm Q}$ satisfies  $|\Delta| \gg g_{\rm Q}$, where $\omega_{\rm C}$ is the cavity frequency, $\omega_{\rm Q}$ is the qubit frequency, and $g_{\rm Q}$ is the qubit-cavity coupling strength, the following dispersive approximation to the Jaynes-Cummings Hamiltonian is realized \cite{Blais04} (with $\hbar = 1$):
\begin{equation}
\hat{H}_{\rm eff} = \left(\omega_{\rm C}+\chi_{\rm Q} \hat \sigma_z \right) \hat a ^\dagger \hat a-\frac{1}{2}(\omega_{\rm Q}-\chi_{\rm Q}) \hat \sigma_z;
\label{eq:dispersive_simple}
\end{equation}
here $\chi_{\rm Q} =g_{\rm Q}^2/\Delta$ is the dispersive coupling strength of the resonator to the qubit, and $\hat \sigma_z$ is the Pauli-\textit{z} operator. One sees from the first term that the effective cavity frequency acquires a shift that depends on the qubit state. It is therefore possible to perform a quantum nondemolition measurement of the qubit by monitoring the microwave transmission across the cavity at a frequency close to the cavity resonance, for example, by using standard homodyne or heterodyne techniques \cite{Blais04,Gambetta07}. This approach for reading out the qubit state through cavity transmission measurements has become standard practice.

Recently much effort has been devoted to the development of near quantum-limited superconducting amplifiers for single-shot detection of the qubit state. Specific milestones include observation of quantum jumps in a transmon qubit \cite{Vijay11}, heralded state preparation of single qubit states to eliminate initialization errors \cite{Johnson12,Riste12}, deterministic preparation of entangled states \cite{Riste13}, stabilization of qubit Rabi oscillations using quantum feedback \cite{Vijay12}, and quantum teleportation \cite{Steffen13}. The technology allows high readout speed \cite{Sank14} and entanglement over large distances \cite{Shankar13}. While this approach works well for a small number of readout channels, the required superconducting amplifiers, cryogenic semiconducting postamplifiers, and quadrature mixers entail significant experimental overhead: the amplifiers often require biasing with a strong auxiliary microwave pump tone which must be isolated from the qubit circuit with bulky cryogenic isolators; moreover, there is no clear path to integrating heterodyne detection at low temperature to provide for a compact, scalable architecture.
\begin{figure}[t]
\begin{center}
\includegraphics[width=.49\textwidth]{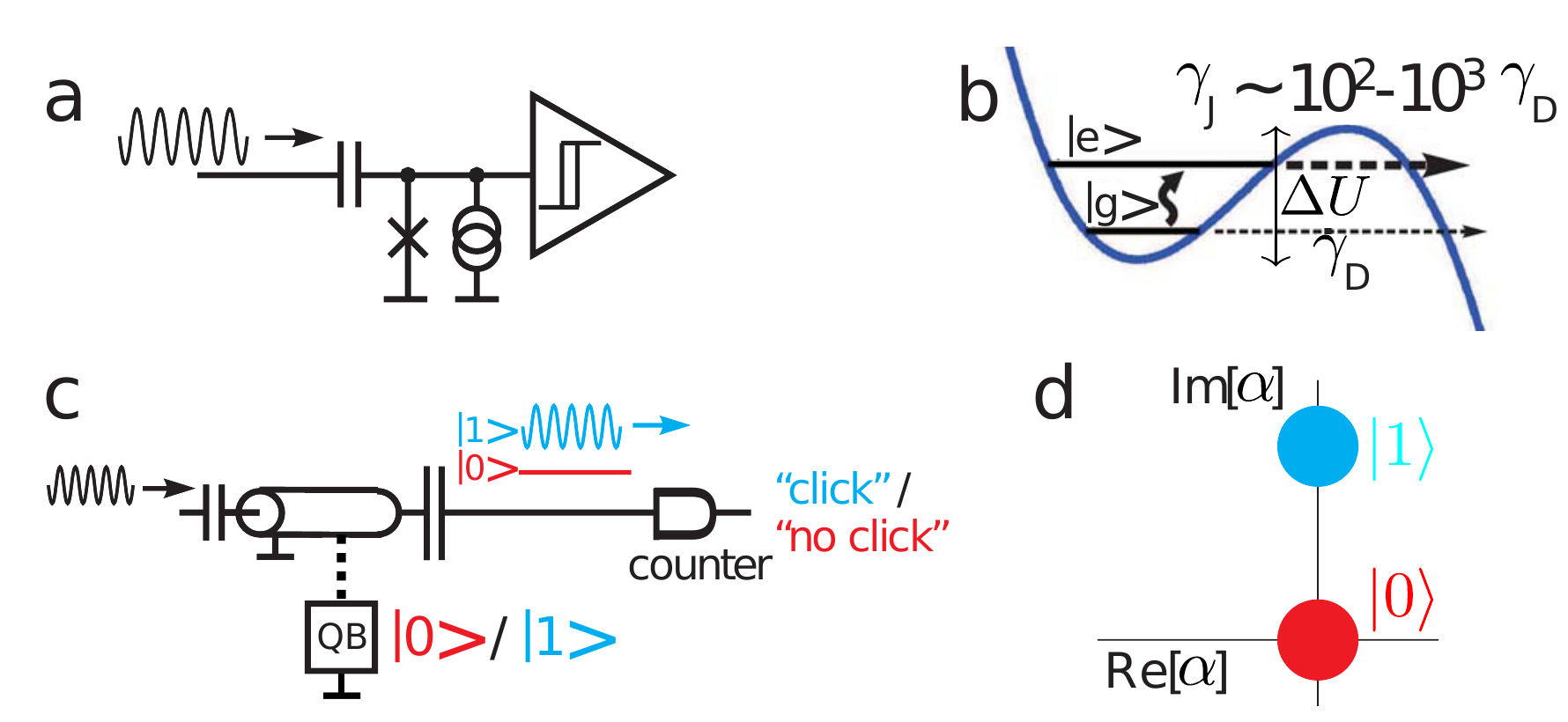}
\vspace*{-0.0in} \caption{(Color online) (a) Schematic diagram of the Josephson photomultiplier (JPM) circuit.  The junction is biased with a dc current, and microwaves are coupled to the junction \textit{via} an on-chip capacitor. In the simplest implementation, switching of the junction creates a voltage pulse that is read out by a room temperature comparator circuit. (b) Junction potential energy landscape. The junction is initialized in the ground state $\ket{g}$. An incident photon induces a transition to the first excited state $\ket{e}$, which rapidly tunnels to the continuum with rate $\yy_{\rm J}$. (c) Counter-based measurement in cQED. ``Bright'' and ``dark'' cavity pointer states result in binary digital output from the JPM: ``click'' or ``no click''. (d) In-phase (I) and quadrature (Q) phase space portrait of the cavity state after the ring-up, highlighting pointers to the $|0\rangle$ state in red and to the $|1\rangle$ state in blue.}
\label{fig:basic}
\end{center}
\end{figure}

An alternative approach that has not yet been considered is to measure the state of the qubit using a photon counter. In contrast to an amplifier, which performs a linear mapping of input modes $\hat{a}, \hat{a}^\dag$ to output modes $\hat{b}, \hat{b}^\dag$, a photon counter responds to the total power of the input signal $\hat{a}^\dag \hat{a}$ in a nonlinear fashion: the presence or absence of photons projects the counter into one of two possible classical output states, irrespective of the phase of the input signal. In the optical frequency range, the prototypical photon counter is the avalanche photodiode \cite{Hadfield:2009lp,Bachor:2004pt}: here, absorption of a single photon creates an electron-hole pair; the reverse bias of the $pn$ junction sweeps the charge away from the depletion region and impact ionization generates additional electron-hole pairs, leading to a large and easily measured classical current.

We have recently introduced a superconducting device that performs as a microwave-frequency analog of the avalanche photodiode \cite{Chen11,Govia:2012uq,Poudel:2012uq}. The detector is a Josephson junction that is biased with a current such that the energy separation between the ground $\ket{g}$ and first excited $\ket{e}$ states in the metastable minima of the junction potential energy landscape is resonant with the energy of the incident microwaves (see Fig. \ref{fig:basic}a-b). Absorption of a single microwave photon promotes the junction from the $\ket{g}$ to the $\ket{e}$ state, which tunnels rapidly to the continuum, producing a large and easily measured voltage of order twice the superconducting gap voltage. We refer to the detector as the Josephson photomultiplier (JPM). The JPM provides an intrinsically broadband frequency response; as we will show here, single-shot measurement contrast around 95\% -- suitable for scalable surface codes \cite{Fowler12} -- is achievable; the detector requires no microwave biasing, facilitating wireup of complex multi-qubit circuits comprising many measurement channels; finally, the detector produces a binary digital output that interfaces well to scalable cold control circuitry based on Single Flux Quantum (SFQ) digital logic \cite{Likharev91}.

This paper is organized as follows. In Section II, we describe the basic principles of the JPM and discuss detector operation. In Section III, we present a detailed theoretical model of the proposed measurement protocol, with a focus on measurement contrast and back action. In Section IV, we discuss how close this scheme comes to a quantum non-demolition (QND) measurement, and in Section V we consider interactions with the environment, taking into account the full Jaynes-Cummings Hamiltonian between the cavity and the qubit. Section VI is devoted to a discussion of issues related to scaling this measurement approach to a large number of readout channels. In Section VII we present our conclusions.

\section*{{\normalsize{\textbf{II. Microwave Photon Counter Based on a Josephson Junction}}}}

\indent \indent A schematic diagram of the JPM is shown in Fig. \ref{fig:basic}a. The Josephson junction is biased in the supercurrent state with a current $I_b$ that is slightly below the junction critical current $I_0$. The potential energy landscape $U(\delta)$ for the phase difference $\delta$ across the junction takes on a tilted-washboard form \cite{Tinkham96}, with local potential minima characterized by a barrier height $\Delta U$ and plasma frequency $\omega_p$ (Fig. \ref{fig:basic}b). The circuit design and bias parameters are chosen so that there is a handful of discrete energy levels in each local minimum of the potential; the JPM initially occupies the ground state $\ket{g}$. Microwaves that are tuned to the junction resonance induce a transition to the first excited state $\ket{e}$, which rapidly tunnels to the continuum. This tunnelling transition in turn leads to the appearance of a large voltage across the junction of order twice the superconducting gap. Absorption of a photon thus yields an unambiguous and easily measured ``click''.

The experimental protocol involves pulsing the bias point of the JPM for a finite interval of order 10s of ns so that the transition frequency between the $\ket{g}$ and $\ket{e}$ states is close to the frequency of the incident photons: at this point, the junction is in the ``active'' state, and there is high probability that absorption of a photon will induce a transition to the continuum. In the absence of resonant photons, there is a small, nonzero probability that the JPM will transition due to quantum tunnelling from $\ket{g}$, a dark-count event. JPM intrinsic contrast peaks for a bias such that $\Delta U / \hbar \omega_p \sim 2$ for a measurement interval that is roughly equal to the Rabi period of the coherent drive  \cite{Poudel:2012uq,Weiss2012}; for very short times, the interaction with the drive field is too weak to induce a transition, while for longer measurement times dark counts due to quantum tunnelling from the ground state degrade performance. In prior work, we have demonstrated efficiencies of order 90\% for coherent drive corresponding to Rabi frequencies around 100 MHz for junctions with extremely modest coherence times of order a few ns \cite{Chen11}.

In the context of qubit measurement, the utility of the JPM is its ability to map bright and dark cavity states to two distinct classical output states: ``click'' or ``no click''. It hence presents a measurement paradigm different from that of a linear amplifier and should be discussed in different terminology \cite{Korotkov:2008ez}. For example, the gain of a JPM at an infinitesimal input signal is negligible as such a signal will not activate it into the voltage state, whereas above a certain threshold the nonlinear gain is extremely high. A performance comparison can, however, be done on the level of the overall qubit measurement protocol.

In a conventional cQED measurement, the state of the qubit is encoded in the quadrature amplitudes of a weak microwave signal that is transmitted across the readout cavity. It is possible to access these amplitudes by preamplifying the signal using a low-noise linear amplifier followed by homodyne or heterodyne detection; assignment of the detected signal to the qubit $\ket{0}$ or $\ket{1}$ states is performed by subsequent post-processing and thresholding. In the following, we analyze an alternative protocol in which the state of the qubit is mapped to the photon occupation of the cavity. The JPM then provides a high-fidelity digital detector of cavity occupation (see Fig. \ref{fig:basic}c-d). The measurement provides no information about the phase of the transmitted microwaves, or indeed about the amplitude of the transmitted signal beyond the digital ``click'' / ``no click'' output of the JPM. As we show below, measurement contrast achievable with the JPM is comparable to that achieved with quantum-limited linear amplifiers, while the JPM provides unique advantages in terms of scaling to a large number of measurement channels. We note that related proposals for photon counters were put forth recently that include both irreversible photon absorption \cite{Romero09,Peropadre11,Andersen13} and non-destructive photon detection via nonlinearity of a transmission line coupled to transmons \cite{Sathya14,Fan14}.

\section*{{\normalsize{\textbf{III. cQED Measurement with a Microwave Photon Counter}}}}

\begin{figure}[t]
%\begin{center}
\includegraphics[width=.5\textwidth]{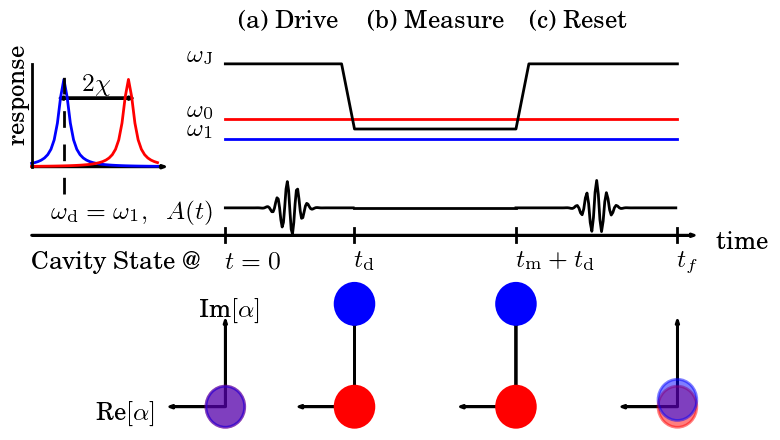}
\vspace*{-0.3in} \hspace*{-4.0in}\caption{(Color online) Schematic of the three-stage measurement protocol: the upper panel describes the relevant control, while the lower panel represents the corresponding cavity state. {\bf (a)} In the drive stage, the cavity is driven strongly and coherently at the cavity frequency dressed by the qubit $\ket{1}$ state,  $\ww_{\rm d} = \ww_{\rm 1} = \ww_{\rm C} - \chi_{\rm Q} + \chi_{\rm J}$,  for a duration $t_{\rm d}=\pi/\chi_{\rm Q}$ (assuming a square pulse).  This projects the qubit onto either  $\ket{0}$ or $\ket{1}$ and conditionally populates the cavity with a large number of photons $n\sim|\alpha_1|^2$ when the qubit is projected onto the $\ket{1}$ state. {\bf (b)} During the measurement phase, the JPM is tuned into resonance with the cavity and allowed to interact;  a bright cavity switches the JPM to its voltage state while a dark cavity leaves the JPM in the supercurrent state. This conditionally squeezes the cavity state by a small amount (not shown here). {\bf (c)} In the reset stage, the cavity is again driven coherently  at $\omega_{\rm d}$, conditionally displacing the cavity to a near-vacuum state.}
\label{fig:JPM_protocol}
%\end{center}
\end{figure}

\indent \indent The basic scheme for qubit measurement with the JPM is shown in Fig. \ref{fig:JPM_protocol}. The qubit (resonating around 5 GHz) is coupled to a readout cavity (resonating around 6 GHz). As in the usual dispersive limit of the Jaynes-Cummings Hamiltonian (\ref{eq:dispersive_simple}), the cavity acquires a dispersive shift $\chi_{\rm Q} \equiv g_{\rm Q}^2/\Delta$ that depends on the state of the qubit. For the purposes of realizing a fast measurement, it is desirable to engineer a dispersive shift $\chi_{\rm Q}/\pi \approx$ 10 MHz, as opposed to the smaller dispersive shifts of order 1 MHz realized in typical cQED experiments. The measurement proceeds in three stages: (1) \textit{Drive stage.} Here, we map the qubit state to microwave photon occupation of the readout cavity. A microwave pulse applied to the dressed frequency corresponding to qubit state $\ket{1}$ creates a ``bright'' cavity if and only if the qubit is in the excited state. If the qubit is in the ground state, the cavity acquires a non-negligible occupation at the start of the pulse, but it coherently oscillates back to the ``dark'' vacuum state upon completion of the drive pulse. During the drive stage the JPM idles at a frequency that is blue detuned from the cavity by around 1 GHz. (2) \textit{Measurement stage.} Here, we map photon occupation of the cavity to the voltage state of the JPM (``click'' or ``no click''). The JPM is rapidly tuned into resonance with the cavity. A bright cavity induces a transition to the voltage state, while a dark cavity leaves the JPM in the supercurrent state. (3) \textit{Reset stage}. It is advantageous to coherently depopulate the bright cavity in order to circumvent the need for the cavity to decay \textit{via} spontaneous emission. However, since the depletion of the cavity due to interaction with the JPM is a stochastic process, so that neither the number of photons removed nor the back action on the cavity is perfectly known or reproducible, it is not possible to return the cavity precisely to the vacuum state. Nevertheless, an appropriate coherent pulse can return the cavity to a state that is close to the vacuum. The measurement pulse sequence is shown in the upper panel of Fig. \ref{fig:JPM_protocol}.

In the dispersive regime  of the qubit-resonator sytem, the unitary evolution of the full system is described by the Hamiltonian
\be
\hat{H} = \hat{H}_{\rm eff} + A(t)\left( \hat{a}+ \hat{a}^{\dagger}\right) - \frac{\ww_{\rm J}(t)}{2}\hat{\sigma}_{z}^{\rm J} + g_{\rm J}\left(\hat{a}\hat{\sigma}_{\rm J}^{+} + \hat{a}^{\dagger}\hat{\sigma}_{\rm J}^{-}\right),
\label{eqn:HamQD}
\ee
where $\ww_{\rm J}(t)$ is the frequency of the JPM, $A(t)$ is the classical drive applied to the cavity, $g_{\rm J}$ is the cavity-JPM coupling, and $\hat{H}_{\rm eff}$ is defined in Eq. (\ref{eq:dispersive_simple}). The JPM operators $\hat{\sigma}_{\rm J}^{\pm}$ couple the ground and excited state of the JPM, which are separated by a frequency $\ww_{\rm J}(t)$ but do not couple to the measured state. The JPM self-Hamiltonian contains $\hat{\sigma}^{\rm J}_{z} = {\rm diag}(1,-1,E)$. Here, the energy of the measured state $E$ is irrelevant once the tunnelling rates (which are not contained in this Hamiltonian as they require interaction with an environment) have been fixed independently.  The measured state plays no role in the unitary dynamics of the system as it only couples incoherently to all other states, and the full dynamics of the JPM are described by a Lindblad-type master equation.

In the following we analyze the three stages of the measurement in detail.

\subsection{Drive Stage}

The goal of this stage is to prepare a photonic state in the cavity that is dependent on the qubit state, such that the conditional cavity states can later be distinguished by the JPM in the measurement stage. The JPM idles in this stage, biased far off-resonance from the cavity such that the effective interaction between the cavity and the JPM is dispersive, with a dispersive shift $\chi_{\rm J} \equiv {g_{\rm J}^2}/{(\ww_{\rm C}-\ww_{\rm J})}$.

The effective Hamiltonian for the cavity becomes
\begin{align}
\hat{H}_{\rm C} =\tilde{\ww}_{\rm C}\hat{a}^{\dagger}\hat{a} + A(t)\left( \hat{a}+ \hat{a}^{\dagger}\right),
\label{eqn:HC}
\end{align}
where $\tilde{\omega}_{\rm C}\equiv  \left(\ww_{\rm C} \pm \chi_{\rm Q} + \chi_{\rm J} \right)$. We choose a classical drive $A(t) = a_0\cos{(\ww_{\rm d} t)}$ for $0\le t\le t_{\rm d}$ where $a_0$ is the drive strength, $\ww_{\rm d}$ the drive frequency, and $t_{\rm d}$ the pulse length (for simplicity here we assume a square pulse). By setting $\ww_{\rm d} = \ww_{\rm C} - \chi_{\rm Q} + \chi_{\rm J}$ we obtain an effective cavity-drive detuning $\delta \ww = \tilde{\ww}_{\rm C} - \ww_{\rm d}$ that depends on the state of the qubit:
\begin{align}
\delta \ww = \left\{
\begin{array}{l l}
2\chi_{\rm Q} &\  \text{for qubit in state $\ket{0}$}\\
0 &\  \text{for qubit in state $\ket{1}$}.
\end{array} \right.\
\end{align}

\begin{figure}
\includegraphics[width = \columnwidth]{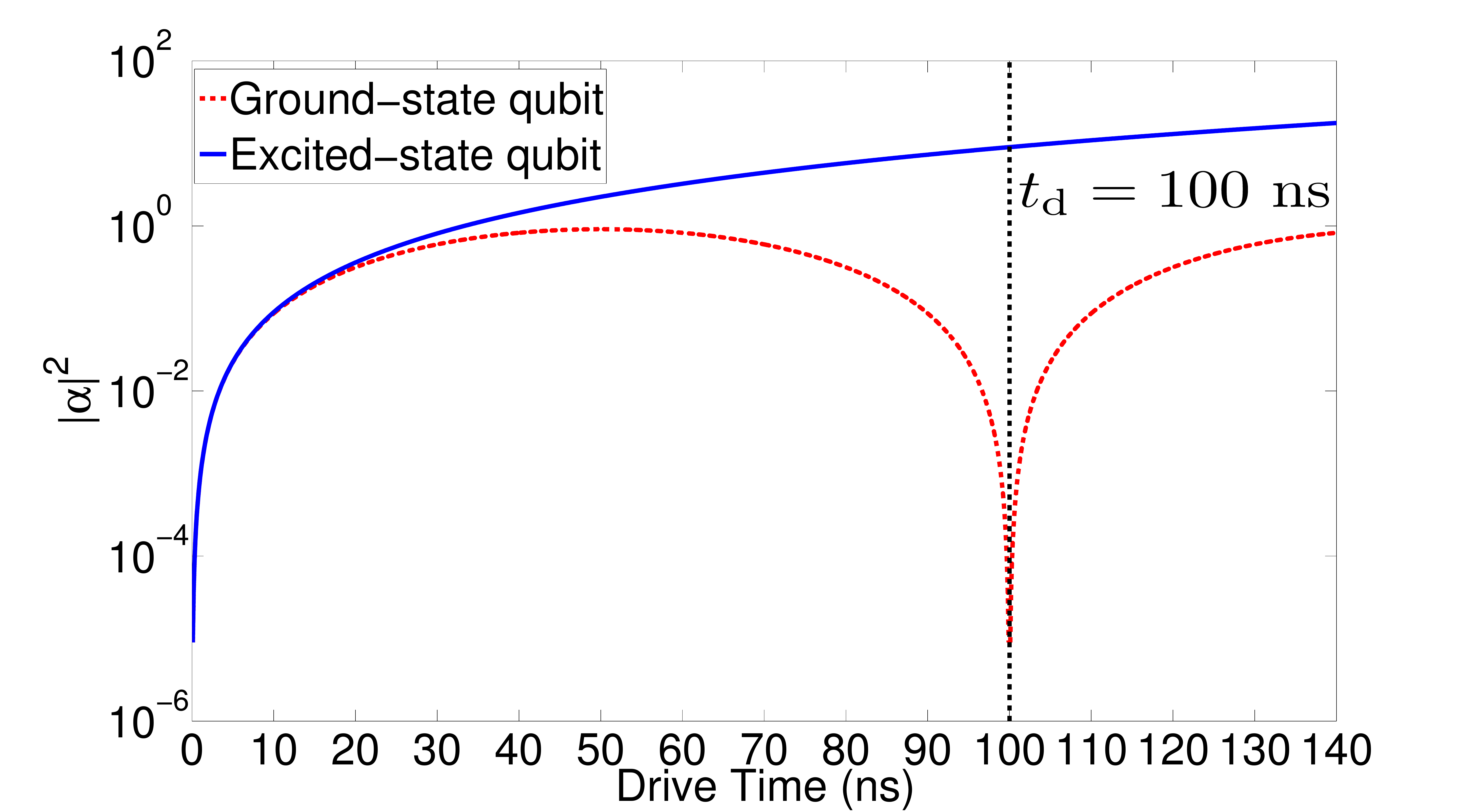}
\caption{(Color online) Cavity occupation as a function of the duration of the applied drive, $t_{\rm d}$, for the qubit in states $\ket{0}$ and $\ket{1}$. Here $\chi_{\rm Q}/\pi = 10$ MHz, such that the optimal drive time is $t_{\rm d} = 100$ ns.}
\label{fig:Drive}
\end{figure}

For such a classical drive of duration $t_{\rm d}$, the solution to Eq. (\ref{eqn:HC}) is easily obtained. Depending on the state of the qubit, the cavity will be in the coherent state $\ket{\alpha_{0/1}}$, with
\begin{align}
\alpha_0 = -\frac{a_0}{4\chi_{\rm Q}}\left( e^{i2\chi_{\rm Q} t_{\rm d}} - 1\right);\ \  \alpha_1 = -\frac{ia_0}{2}t_{\rm d}
\label{eqn:als1}
\end{align}
up to a global phase.
We see that when the qubit is in state $\ket{0}$, the cavity occupation oscillates sinusoidally at a frequency set by the detuning of the drive pulse from the  dressed cavity resonance. On the other hand, when the qubit is in state $\ket{1}$, the cavity occupation $|\alpha_1|^2$ grows monotonically in time. This is shown in Fig. \ref{fig:Drive}, where we plot cavity occupation \textit{versus} coherent drive time for the qubit in states $\ket{0}$ and $\ket{1}$.
In order to maximize contrast between the dark and bright cavity states to which the qubit states $\ket{0}$ and $\ket{1}$ are mapped, it is optimal to choose $t_{\rm d} = \pi/\chi_{\rm Q}$ such that $\alpha_0 = 0$ at the end of the drive stage. The length of the drive stage is therefore set by the requirement that $\alpha_0(t_{\rm d})= 0$ and not by the input cavity coupling, which is the inverse of the decay time of the cavity through its input port.

We assume the system starts in the state
\be
\label{eqn:start}
\ket{\Psi(0)} = \ket{0}_{\rm C}\otimes\ket{\psi}_{\rm Q}\otimes\ket{0}_{\rm J},
\ee
where the qubit state $\ket{\psi}_{\rm Q} = a\ket{0} + b\ket{1}$ can be prepared independently by the qubit drive line. After the drive stage, the system is left in the state
\be
\ket{\Psi(t_{\rm d})} = \big(a\ket{\alpha_0}_{\rm C}\otimes\ket{0}_{\rm Q} + b\ket{\alpha_1}_{\rm C}\otimes\ket{1}_{\rm Q} \big)\otimes\ket{0}_{\rm J},
\label{eqn:DriveOut}
\ee
which can be verified by solving Eq. (\ref{eqn:HamQD}) analytically. In the case that $\alpha_0 = 0$, the cavity has nonzero occupation only when the qubit is in the excited state.

The drive stage can be thought of as the first step in a quantum measurement of the qubit state, as described in the pointer basis formalism of Zurek \cite{Zurek:1981vq}. In this language, the cavity states $\ket{\alpha_{0/1}}$ form the pointer basis that is entangled with the qubit. Examining the reduced density matrix of the qubit state
\begin{align}
\hat{\rho}_{\rm Q} = \left(
\begin{array}{cc}
\abs{a}^2 & a^*bD \\
ab^*D & \abs{b}^2
\end{array}
\right),
\end{align}
we see that qubit coherence has been suppressed by a factor $D = {\rm exp}\left(-\abs{\alpha_1 -\alpha_0}^2\right)$, which quantifies the dephasing of the qubit induced by the interaction with the pointer basis (cavity). The dephasing would be complete if the pointer states were orthogonal. Moreover, mapping of the qubit $\ket{1}$/$\ket{0}$ states to bright/dark cavity states can be viewed as a coherent amplification step, as the information about the qubit state is now contained in a large number of photons. A more detailed discussion of the consequences of this overlap on the detection contrast and back action will be presented later.

As a result of the strong dephasing of the qubit state during the drive stage (quantified by the factor $D$), our multi-stage protocol explicitly exposes the role of the pre-measurement stage in quantum non-demolition (QND) readout. In particular, our protocol highlights the fact that in QND readout of the qubit state, measurement of the cavity pointer states is not the major source of qubit state dephasing. The qubit states are dephased during the pre-measurement, when qubit states and cavity pointer states are entangled, which in our case corresponds to the drive stage. The main role of the subsequent pointer state measurement (the measurement stage in our protocol) is to break unitarity and ``freeze'' the qubit in a dephased state. This distinction between pre-measurement and measurement is less obvious in qubit readout using a continuous cavity signal with linear amplification and heterodyne detection. The clear distinction between pre-measurement and measurement in our protocol allows for independent control of each stage, which can be used to achieve higher readout fidelity (as has been done here), and to study, both in theory and experiment, QND measurement and the pointer basis formalism with an explicit physical system in mind. A similar distinction between pre-measurement and measurement exists in a readout scheme in atomic cavity QED, albeit in a rather different parameter regime. This scheme employs dispersive coupling between the cavity and a travelling atom (pre-measurement) followed by atomic state detection via ionization (measurement) to read out the cavity state \cite{Bertet:2002fk,Gleyzes:2007jk}.

\subsection{Measurement Stage}

After the drive stage, the qubit state information has been transferred to the cavity occupation. In the measurement stage, a measurement of the cavity by the JPM will reveal the state of the qubit. During this stage, the JPM is brought into resonance with the dressed frequency of the cavity corresponding to the qubit $\ket{1}$ state, $\ww_{\rm J} = \ww_{\rm C} - \chi_{\rm Q}$, in order to maximize detection in the case that the qubit is excited. In practice, precise tuning of the JPM bias point is not required due to the broad detection bandwidth of the JPM \cite{Poudel:2012uq}.

The Hamiltonian during this stage is that of Eq. (\ref{eqn:HamQD}) with $A(t) =0$. In the following, we assume a cavity-JPM coupling $g_{\rm J}/2\pi = 50$ MHz. In addition, the system evolves incoherently as a result of tunnelling (both bright and dark) and relaxation of the JPM. We consider tunnelling from both the JPM excited and ground states to the measured state, and relaxation from the excited state to the ground state, with corresponding rates $\gamma_{\rm J}$, $\gamma_{\rm D}$, and $\yy_{\rm R}$, respectively. Here we take $\gamma_{\rm J}$ = 200 MHz, $\gamma_{\rm D}$ = 1 MHz, and $\yy_{\rm R}$ = 200 MHz; this relaxation rate corresponds to a junction with capacitance 100 pF directly connected to an environmental impedance of 50 $\Omega$. The total evolution of the system can therefore be described by a Lindblad-type master equation with Lindblad operators corresponding to each incoherent process of the JPM, as outlined in more detail in our previous work \cite{Govia:2012uq,Poudel:2012uq}.

As the cavity-JPM coupling and bright count rate can be controlled independently of one another, they can be adjusted into an optimal regime for good measurement. As explained in more detail in our previous work \cite{Chen11,Poudel:2012uq}, the optimal regime for good measurement is when $g_{\rm J} \sim  \gamma_{\rm J}$, as in this regime the bright count rate is large enough for a bright count to occur within the occupation time of the JPM (per Rabi cycle), while not so large as to result in a Zeno effect suppression of the cavity-JPM interaction. The coupling and bright count rate chosen for the numerical simulations presented here are well within the optimal regime for good measurement.

Starting from the output state of the drive stage, we numerically solve the master equation for the measurement stage to obtain the detection probability $P\left(\abs{\al}^2,t_{\rm m}\right)$ as a function of cavity occupation and measurement time $t_{\rm m}$. In the case that the qubit starts in the $\ket{0}$/$\ket{1}$ state, the detection probability reduces to $P\left(\abs{\al_{0/1}}^2,t_{\rm m}\right)$. We define the qubit measurement contrast as the difference in detection probability between these two cases:
\be
C = P\left(\abs{\al_1}^2,t_{\rm m}\right) - P\left(\abs{\al_0}^2,t_{\rm m}\right).
\label{eq:contrast_def}
\ee
Clearly, the measurement contrast is optimized when $P\left(\abs{\al_0}^2,t_{\rm m}\right) = 0$, which requires that $\al_0 = 0$ and $\yy_{\rm D} = 0$. The measurement contrast has a maximal value of one if $|\alpha_1|\rightarrow\infty$, indicating a perfect measurement.
\begin{figure}
\subfigure{
\label{fig:ProbPT}
\includegraphics[width = \columnwidth]{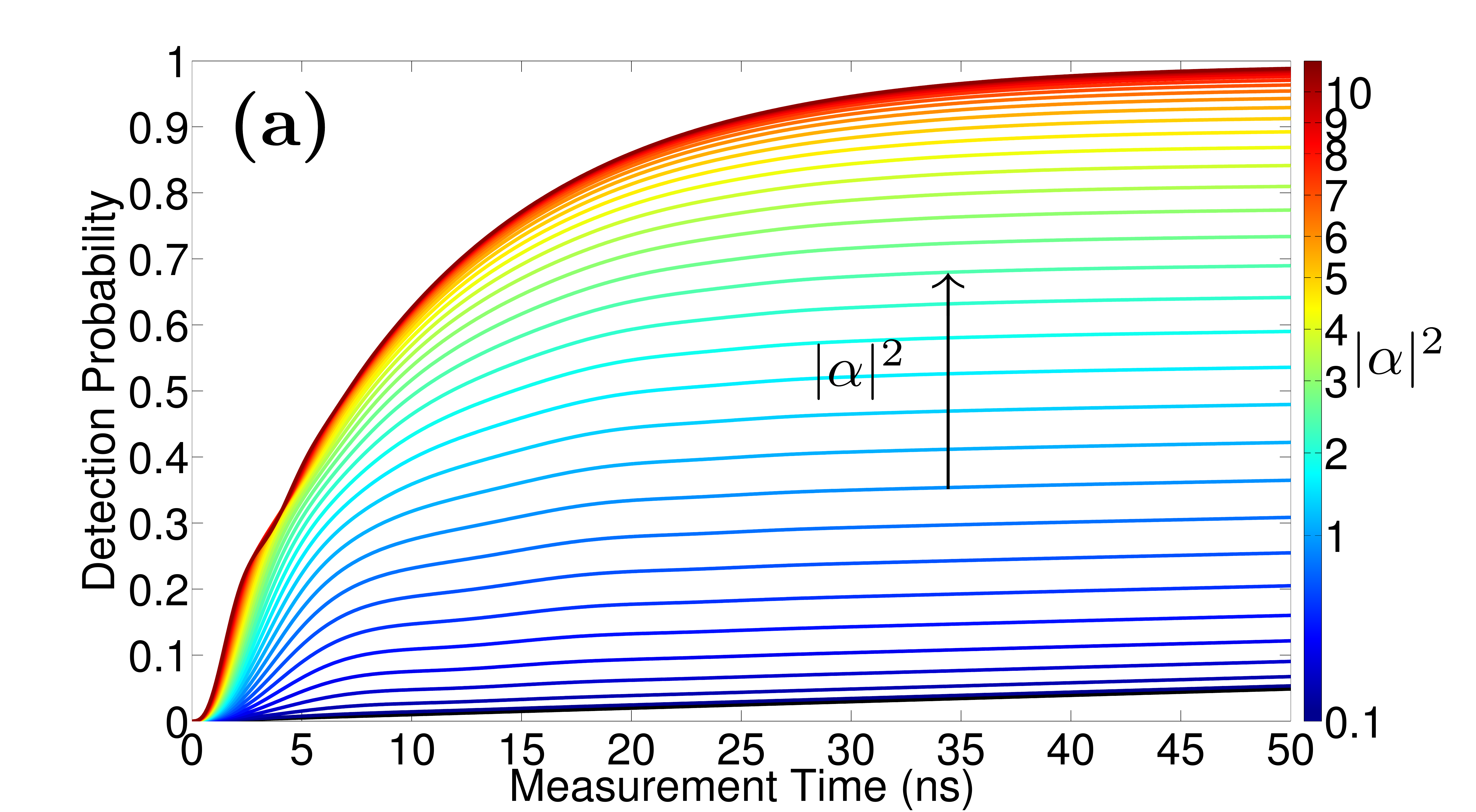}}
\subfigure{
\label{fig:ProbP25}
\includegraphics[width = \columnwidth]
{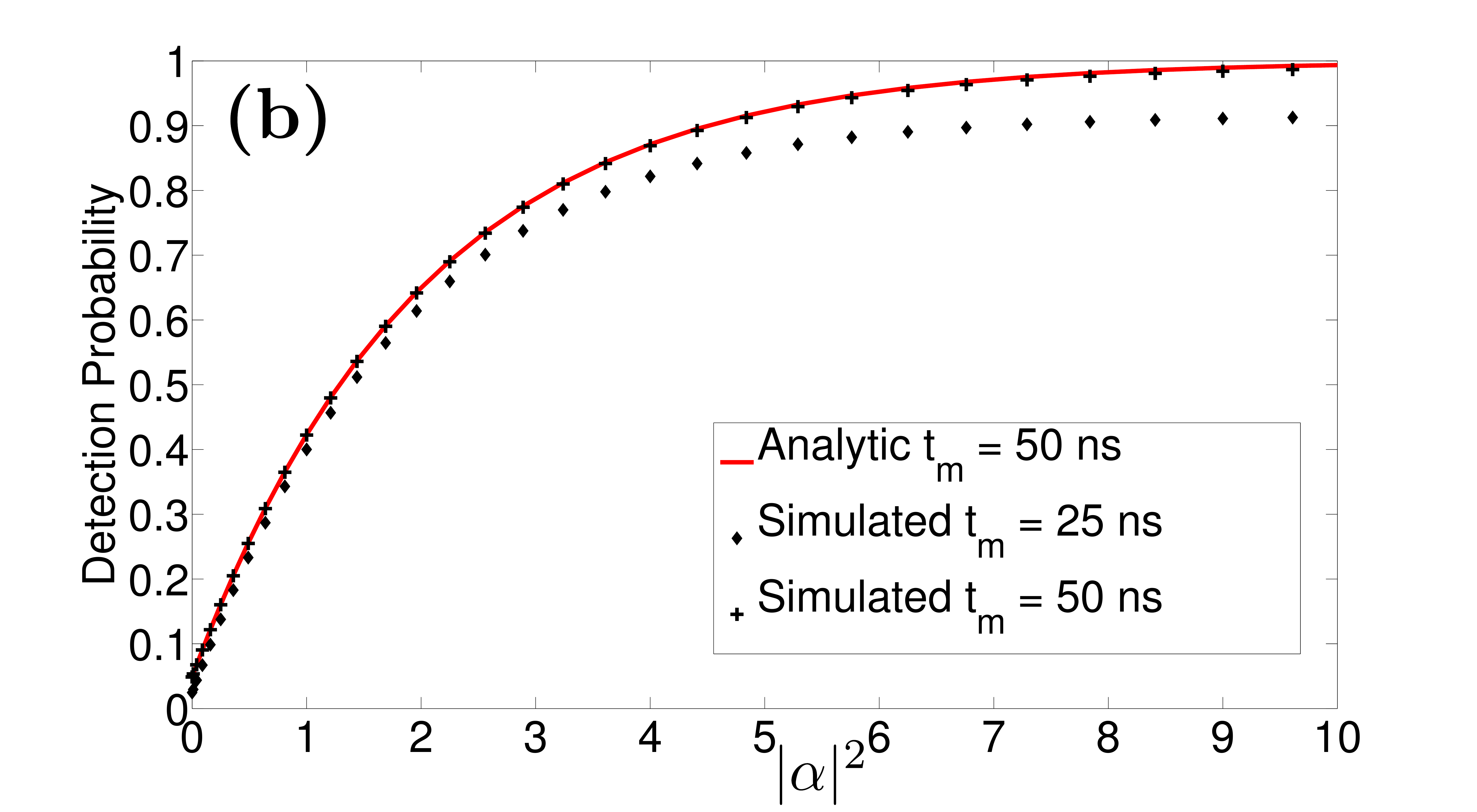}}
\caption{(Color online) {\bf (a)} Detection probability $P\left(\abs{\al}^2,t_{\rm m}\right)$ as a function of measurement time $t_{\rm m}$ (horizontal axis) and cavity occupation $\abs{\al}^2$ (color) for system parameters $g_{\rm J}/2\pi  = 50$ MHz, $\yy_{\rm J}  = 200$ MHz, $\yy_{\rm D}  = 1$ MHz and $\yy_{\rm R}  = 200$ MHz.
{\bf (b)} Detection probability $P\left(\abs{\al}^2,t_{\rm m}\right)$ as a function of cavity occupation $\abs{\al}^2$, both numerical simulations and analytic fit, for $t_{\rm m} = 25$ and 50 ns.}
\end{figure}

Figure \ref{fig:ProbPT} shows the detection probability as a function of both $\abs{\al}^2$ and $t_{\rm m}$ for the system parameters previously discussed. For all coherent states with average photon number $\abs{\al}^2 > 0$ we see similar behavior as a function of time, with saturation of the detection probability around 40 ns, irrespective of cavity occupation.
The fact that the detection probability saturates at a value less than unity is explained by the two competing mechanisms for excitation loss in the JPM: measurement tunnelling and inelastic relaxation, only the former of which results in a bright count. The black curve in Fig. \ref{fig:ProbPT}, $\al = 0$, is the detection probability for the cavity when the qubit is in the ground state, $P\left(\abs{\al_{0}}^2 = 0,t_{\rm m}\right)$, and the fact that it is nonzero is due only to dark counts, which occur with a probability $P_{\rm D}(0,t) = 1 - e^{-\gamma_{\rm D} t}$.

In a simplified picture where energy in the detector automatically leads to a click, we would have
\be
\label{eqn:PDC}
P(\abs{\al}^2) = P_{\rm B}(\abs{\al}^2) + (1 - P_{\rm B}(\abs{\al}^2))P_{\rm D},
\ee
where $P_{\rm B}(\abs{\al}^2)$ is the bright count probability and $P_{\rm D} = P_{\rm D}(0,t)$ is the dark count probability, which we take to be independent of $\al$. However, the detection mechanism of a JPM involves the coherent absorption of energy prior to a tunnelling transition to the classically observable voltage state. In the interval following absorption of a photon but prior to tunnelling, dark counts cannot occur as the JPM is in its excited state. This breaks the dark/bright symmetry of Eq. (\ref{eqn:PDC}); as a result, this equation may no longer be valid. However, as it shows how dark counts are detrimental to measurement contrast, Eq. (\ref{eqn:PDC}) is a good reference point to compare against, and will still be valid in some situations, such as when the JPM coupling rate is smaller than the bright tunnelling rate, i.e., $g_{\rm J} < \yy_{\rm J}$.

Figure \ref{fig:ProbP25} shows the detection probability as a function of $\abs{\al}^2$ for $t_{\rm m} = 25$ and 50 ns, along with an analytic fit  to the data by Eq. (\ref{eqn:PDC}), with $P_{\rm B}$ given by the curve
\be
\label{eqn:PB}
P_{\rm B}(\abs{\al}^2,t_{\rm m} \rightarrow \infty) = 1 - \exp\left(-|\alpha|^2 \frac{\yy_{\rm J}}{\yy_{\rm J}+\yy_{\rm R}}\right).
\ee
See Appendix \ref{app:BR} for a derivation. The analytic curves for both $t_{\rm m} = 25$ and 50 ns are so similar on this scale that only $t_{\rm m} = 50$ ns is plotted. As can be seen, the analytic fit is valid when $t_{\rm m}$ is sufficiently large. For small $t_{\rm m}$, Eq. (\ref{eqn:PDC}) remains close to correct, but the approximation for $P_{\rm B}$ in Eq. (\ref{eqn:PB}) breaks down.

\begin{figure}
\subfigure{
\label{fig:ProbBDC}
\includegraphics[width = \columnwidth]{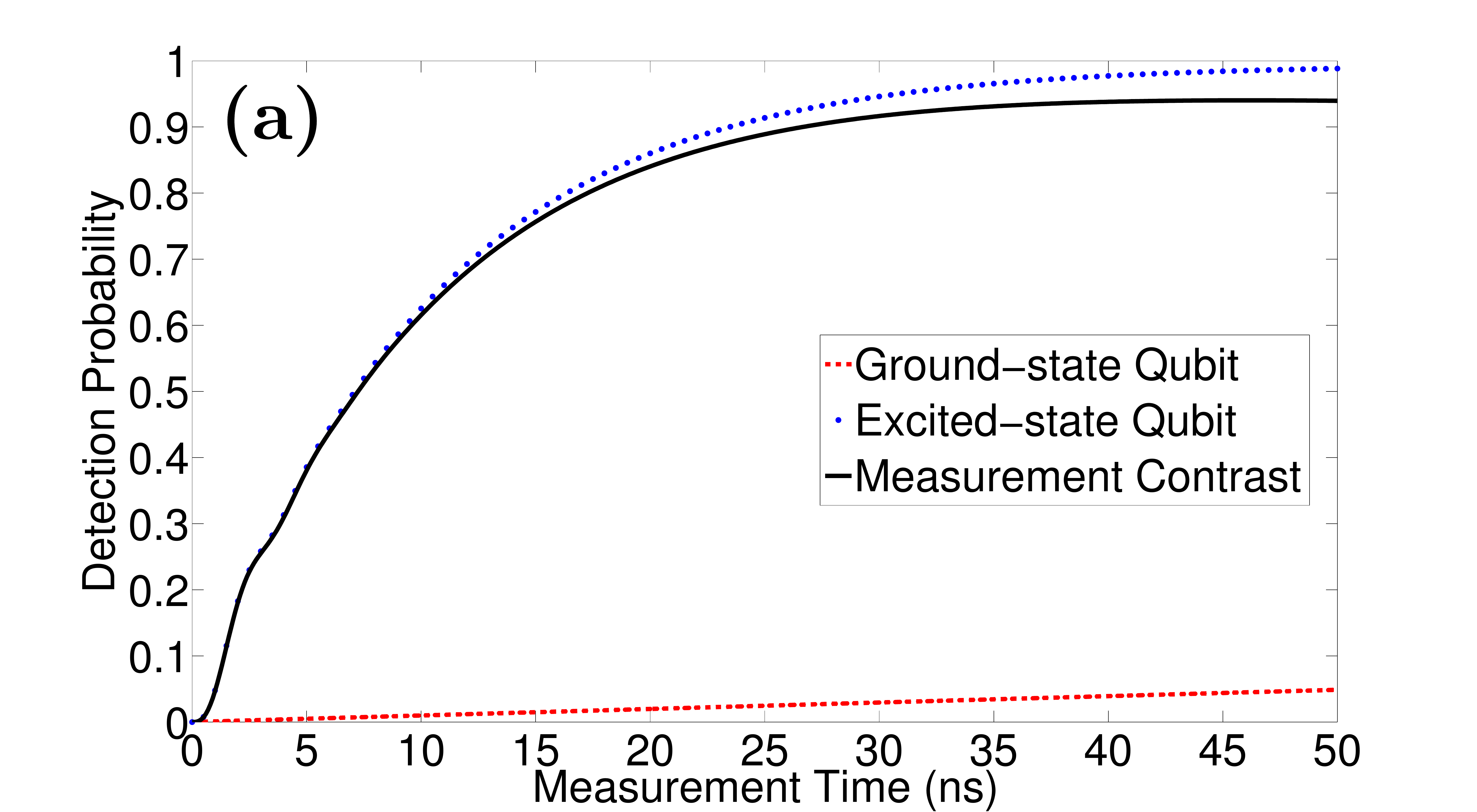}}
\subfigure{
\label{fig:Csurf}
\includegraphics[width = \columnwidth]{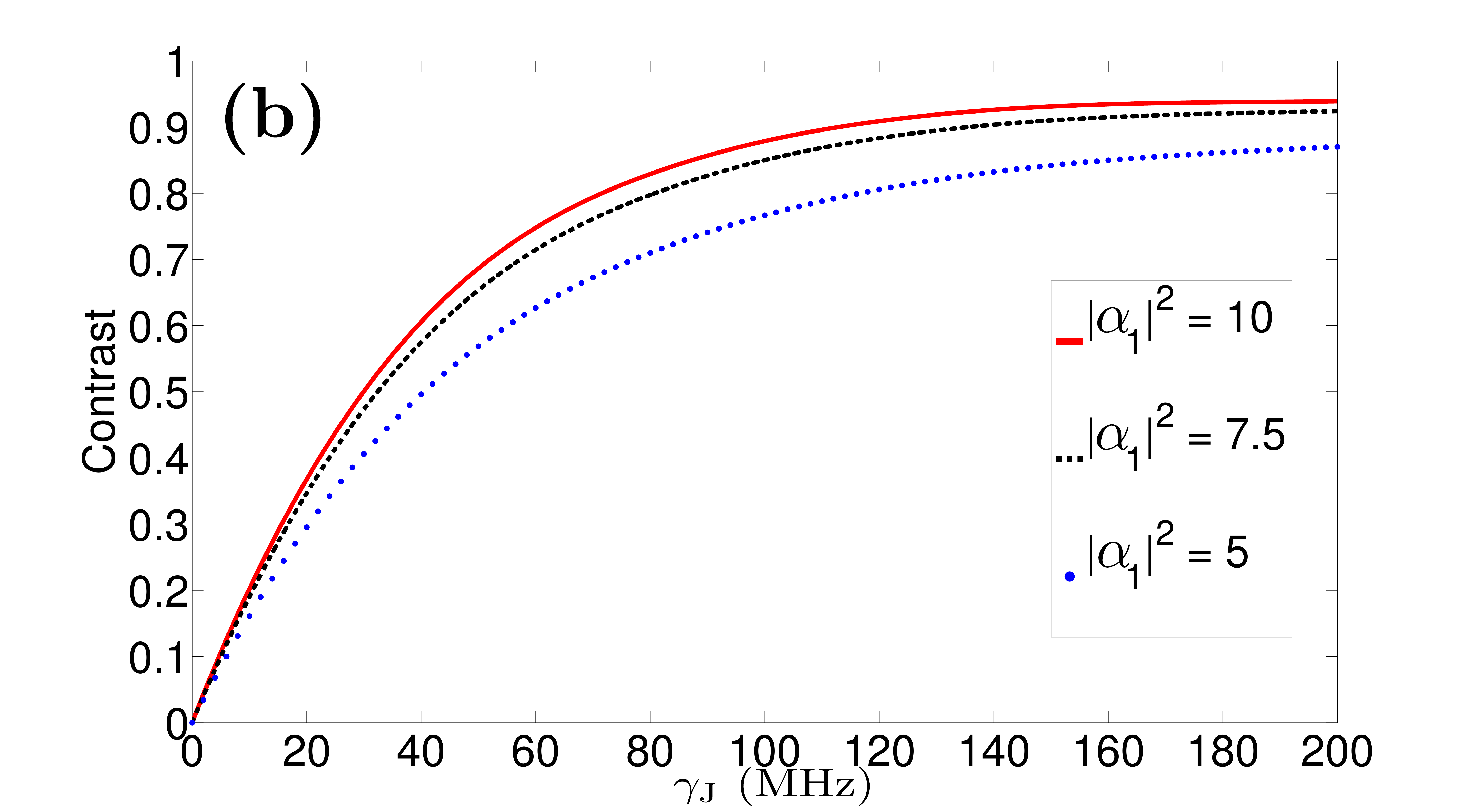}}
\caption{(Color online) {\bf (a) }Excited qubit detection probability $P\left(\abs{\al_{1}}^2,t_{\rm m}\right)$, ground qubit detection probability $P\left(\abs{\al_{0}}^2,t_{\rm m}\right)$, and measurement contrast \textit{versus} measurement time. Here $\abs{\al_{1}}^2 = 10 $ and $\abs{\al_{0}}^2 = 0$, with system parameters as before. {\bf (b)} Measurement contrast as a function of bright count rate $\yy_{\rm J}$, for various $\abs{\al_{1}}^2$ and $t_{\rm m} = 50$ ns. The relaxation rate remains fixed at $\yy_{\rm R} = 200$ MHz, while the dark count rate changes such that the ratio $\yy_{\rm J}/\yy_{\rm D} = 200$ is unchanged.}
\end{figure}

We have calculated detection probability $P\left(\abs{\al}^2,t_{\rm m}\right)$ and measurement contrast \textit{versus} measurement time $t_{\rm m}$ for $\abs{\al_0}^2 = 0$, and $\abs{\al_1}^2 = 10$; the results are shown in Fig. \ref{fig:ProbBDC}. The measurement contrast peaks at $\approx 95 \%$ around 40 ns, indicating that a good choice for $t_{\rm m}$ is 40~ns. At longer times the measurement contrast will eventually begin to decrease, as $P\left(\abs{\al_{0}}^2,t_{\rm m}\right)$ continues to increase while $P\left(\abs{\al_{1}}^2,t_{\rm m}\right)$ asymptotes to near unity. In general, we observe that increasing $\alpha_1$ increases the contrast, ultimately limited by the breakdown of the dispersive approximation to the Jaynes-Cummings Hamiltonian.

The contrast shown in Fig. \ref{fig:ProbBDC} is for one set of system parameters, and in principle it is possible to obtain higher values of measurement contrast by optimizing over parameter space. Figure \ref{fig:Csurf} shows the optimal measurement contrast as a function of bright count rate, $\yy_{\rm J}$, for various bright states $\abs{\al_1}^2$. The ratio of the bright and dark count rates is set by fabrication parameters of the JPM, and therefore remains fixed at $\yy_{\rm J}/\yy_{\rm D} = 200$. However, the inelastic relaxation rate remains fixed as $\yy_{\rm J}$ changes, such that $\yy_{\rm R} = 200$ MHz regardless of the value of $\yy_{\rm J}$. As can be seen in Fig. \ref{fig:Csurf}, within experimentally reachable parameter regimes contrast greater than 95\% is possible.

Measurement contrast is ultimately limited by the possibility of misidentifying the qubit state. Misidentification of the excited state as the ground state is due to the nonzero vacuum component of the coherent state $\ket{\al_1}$ as well as to internal photon loss. This occurs with a probability $1 - P\left(\abs{\al_1}^2,t_{\rm m}\right)$, which is bounded below by $e^{-\abs{\al_1}^2}$ (occuring when $\yy_{\rm R} = 0$). Misidentification of the ground state as the excited state is the result of a dark count (assuming $\al_0 = 0$), and the probability of misidentification in this case is exactly $P_{\rm D}(0,t_{\rm m})$ discussed earlier. The problem of misidentification, and the fact that measurement contrast is less than unity even for $\yy_{\rm R}, \yy_{\rm D} = 0$, is related to the basis of our measurement protocol and will be discussed in more detail in Section IV.

After the measurement stage, if the JPM absorbs a photon and switches out of the supercurrent state, classical emission due to this switching process could induce relaxation in the qubit or produce a spurious population in the readout cavity that would spoil the reset pulse \cite{McDermott05}. The resulting population is proportional to the energy spectral density of the classical current drive at the qubit or cavity frequency. A straightforward approach to address this would be to install a microwave isolator between the cavity and JPM, as in conventional cQED experiments, where one inserts one or more cryogenic isolators between the measurement cavity and the superconducting preamplifier. The breaking of time-reversal symmetry by the isolator allows signals to travel from the cavity to the readout device with minimal loss, while back action noise is heavily attenuated. However, we anticipate that this classical back action can be greatly suppressed by an appropriate choice of JPM parameters to suppress harmonics of the switching transients at the qubit and cavity frequencies, or by shunting the JPM by an appropriate admittance to prevent a full switch of the JPM phase to the running state.  Alternatively, it might also be possible to eliminate a cryogenic isolator by incorporating on the JPM chip tunable impedance-matching circuitry, as this would allow for the realization of a strong impedance mismatch between the cavity output and JPM immediately after the end of the measurement stage.

\subsection{Reset Stage}

The final stage is to remove the energy from the cavity, ideally leaving the cavity-qubit system in the conditional states $\ket{0}_{\rm C}\ket{0}_{\rm Q}$ or $\ket{0}_{\rm C}\ket{1}_{\rm Q}$ to allow for additional operations on the qubit. This can be achieved through cavity decay by simply waiting long enough; however, because the total cavity decay time may be comparable to the qubit $T_1$, it is preferable to actively reset the cavity.

After the measurement stage, the cavity is either in the vacuum state or the state $\hat{B}_{\rm J}\hat{B}_1^{N-1}\ket{\al_1}$. Here, $\hat{B}_{{\rm J},1}$ are the back action operators \cite{Govia:2012uq} on the cavity due to JPM tunnelling and inelastic relaxation, respectively, and $N$ is the number of photons removed from the cavity by the JPM. These back action operators interpolate between the standard lowering operator $\hat{B}=\hat{a}$ and the subtraction operator  $\hat{B}=\hat{a}\hat{n}^{-1/2}$ \  \cite{Govia:2012uq}.
We neglect for the moment the classical back action on the cavity due to the transient current that develops when the JPM switches to the voltage state. As a starting point for reset, we will assume that $\hat{B}_{\rm J}\hat{B}_1^{N-1}\ket{\al_1} \approx \ket{\al_{\rm M}}$ even with large $\gamma_{\rm R}$, with
\be
\abs{\al_{\rm M}}^2 = {\rm Tr}\left[\hat{a}^{\dagger}\hat{a}\hat{B}_{\rm J}\hat{B}_1^{N-1}\ket{\al_1}\bra{\al_1} \hat{B}_{\rm J}^{\dagger}\hat{B}_1^{\dagger N-1}\right],
\ee
the average photon number of the cavity state after measurement.

At the end of the reset stage, we desire the cavity to be in the vacuum state independent of the qubit state, and thus we must invert the drive stage. Consider a Hamiltonian of the form of Eq. (\ref{eqn:HC}), with a more general drive $A(t) = a_1\cos{(\ww_{\rm d} t+\phi)}\Theta(t_{\rm d}-t)$, where $t_{\rm d} = \pi/\chi_{\rm Q}$ as before. The unitary operation applied to the cavity is then
\be
\hat{U}_{\rm r} = \mathbb{I}_{\rm C}\otimes\ket{0}\bra{0}_{\rm Q} + \hat{D}(\beta)\otimes\ket{1}\bra{1}_{\rm Q}.
\ee
Here $\hat{D}(\beta)$ is the displacement operator on the cavity, with
$$
\beta = \frac{-ia_1t_{\rm d}}{2}e^{-i\phi}.
$$
Thus, by choosing $a_1$ such that $(a_1t_{\rm d})/2 = \abs{\al_{\rm M}}$ and setting $\phi = (2n+1)\pi, n \in \mathbb{Z}$, we have $\beta = -\al_{\rm M}$. Under these conditions, the operation $\hat{U}_{\rm r}$ will leave the cavity in the vacuum state independent of the qubit state, and will do so with an operation time $t_{\rm r} = t_{\rm d}$, significantly shorter than the total decay time of the cavity.

\begin{figure}
\includegraphics[width = \columnwidth]{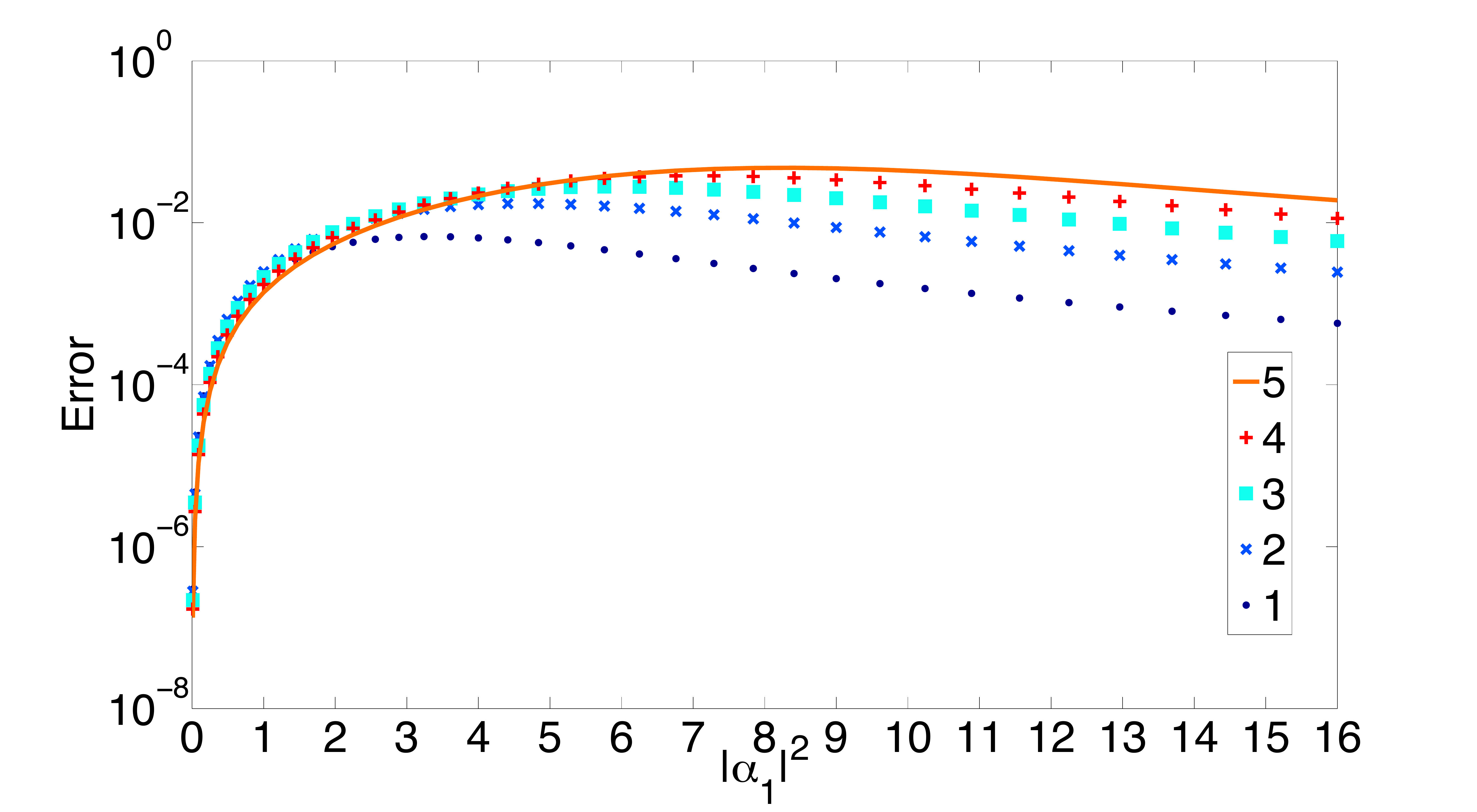}
\caption{(Color online) Numerically evaluated overlap error for the reset pulse described in the text. The number of photons $N$ removed from the cavity is shown in the legend.}
\label{fig:Evac}
\end{figure}

However, after detection by a JPM the state of the cavity is not a coherent state; thus there does not exist a displacement operator $\hat{D}(\beta)$ that will map it identically to the vacuum state. One can quantify the resulting deviation from vacuum by calculating
\be
E(\al_1,N) = 1- \frac{1}{A^2}\abs{\bra{0}D(-\al_M)\hat{B}_{\rm J}\hat{B}_1^{N-1}\ket{\al_1}}^2,
\ee
where $A$ is the normalization of the state after measurement. This error will depend on the form of the back action. Assuming all back actions can be expressed in terms of subtraction operators as in [\onlinecite{Govia:2012uq}], we find
\begin{align}
\nonumber &E(\al_1,N) = 1- \frac{1}{P_N}\abs{\bra{-\al_M}\hat{B}_{-}^{N}\ket{\al_1}}^2 \\
&= 1 - \frac{1}{P_N}\left|e^{-\frac{\abs{\al_M}^2+\abs{\al_1}^2}{2}}\sum_{n=0}^{\infty}\frac{\al_M^n\al_1^{n+N}}{\sqrt{n!(n+N)!}}\right|^2.
\label{eqn:E}
\end{align}
Here, the normalization $A^2$ is the probability of $N$ photons being subtracted \cite{Govia14}, $P_N = 1 - \frac{\Gamma(N,\abs{\al_1}^2)}{\Gamma(N)}$, where $\Gamma(N,\abs{\al_1}^2)$ is the upper incomplete Gamma function. The error of this reset pulse is shown in Fig. \ref{fig:Evac} for different values of $N$ and as a function of $\abs{\al_1}^2$. As can be seen, the maximal error increases with increasing $N$, but for all $N$ the error tends to zero as $\abs{\al_1} \rightarrow \infty$. In reality, as the value of $N$ is not fixed, a better estimate for the average error can be obtained by averaging over the error traces shown in Fig. \ref{fig:Evac}.
Note that if the back action operator is closer to the standard photon lowering operator $\hat{a}$, this figure of merit will improve as the coherent states are eigenstates of this operator, and can be moved to vacuum exactly.

The possibility exists that more complicated pulse sequences during the reset phase will be able to map the cavity state identically to the vacuum; however, consideration of such sequences is beyond the scope of this work. In any case, the error of the reset pulse shown here does not affect the success of qubit readout.

\section*{{\normalsize{\textbf{IV. QNDness of the Measurement}}}}
\label{sec:QND}

Ideally, we would like our measurement to project the qubit into its bare basis, $\{\ket{0/1}_Q\}$, hence implementing a quantum nondemolition (QND) measurement. A hallmark of QND measurement is that a repeated measurement leads to the same result with certainty. Our measurement scheme starts from the dispersive Hamiltonian of Eq. (\ref{eq:dispersive_simple}) in the cavity ring-up phase, which is QND in the sense that the qubit and the pointer coupling commute. However, the process of destructive photon absorption necessarily results in  a deviation from QNDness, which we analyze in detail below. Even in the case of an ideal measurement ($\yy_{\rm R}, \yy_{\rm D} = 0$ and $t_{\rm m} \rightarrow \infty$), the potential misidentification of the two states leads to a QND error. Starting from the state in Eq. (\ref{eqn:DriveOut}), the measurement  projects the qubit conditionally onto the states
\begin{align}
\label{eqn:psi0}
&\ket{\psi_0} = \frac{ae^{-\abs{\al_0}^2/2}\ket{0}_{\rm Q} + be^{-\abs{\al_1}^2/2}\ket{1}_{\rm Q}}{\sqrt{\abs{a}^2e^{-\abs{\al_0}^2} + \abs{b}^2e^{-\abs{\al_1}^2}}} \\
\label{eqn:psi1}
&\ket{\psi_1} = \frac{a\sqrt{1-e^{-\abs{\al_0}^2}}\ket{0}_{\rm Q} + b\sqrt{1-e^{-\abs{\al_1}^2}}\ket{1}_{\rm Q}}{\sqrt{\abs{a}^2(1-e^{-\abs{\al_0}^2}) + \abs{b}^2(1-e^{-\abs{\al_1}^2})}}.
\end{align}
Even for $\al_0=0$ these states are non-orthogonal (and not equal to the ideal QND post-measurement states), and their overlap is related to the overlap of the cavity pointer states $\ket{\al_{0/1}}$. This overlap is what allows misidentification to occur, ultimately limiting the contrast and QNDness.

If we do not condition on the measurement outcome, the effect of a perfect QND measurement is the quantum process defined by the map
\be
a\ket{0} + b\ket{1} \longrightarrow \abs{a}^2\ket{0}\bra{0} + \abs{b}^2\ket{1}\bra{1},
\ee
which completely destroys all coherence in the qubit state, while maintaining relative populations. We can describe this quantum process by its Choi matrix $\hat{\mathcal{C}}_{\rm per}$ (see Appendix \ref{app:Fid}), and can compare this to the Choi matrix $\hat{\mathcal{C}}_{t_{\rm m}}$ describing our measurement protocol (which is a function of the measurement time $t_{\rm m}$) using the following Choi matrix fidelity
\be
\mathcal{F}_{t_{\rm m}} = \frac{\left({\rm Tr}\left[\sqrt{\sqrt{\hat{\mathcal{C}}_{\rm per}}\hat{\mathcal{C}}_{t_{\rm m}}\sqrt{\hat{\mathcal{C}}_{\rm per}}} \right]\right)^2}{{\rm Tr}\left[\hat{\mathcal{C}}_{\rm per}\right]{\rm Tr}\left[\hat{\mathcal{C}}_{t_{\rm m}}\right]}.
\label{eqn:fid}
\ee
As this fidelity compares the unconditional measurement protocol, it does not contain information about the success of the measurement (which we believe is well described by the contrast), but instead quantifies how close the possible qubit output states are to the desired ones. As a result, by choosing ideal QND measurement as our reference process, we can directly quantify the QNDness of our measurement protocol.

As we are examining the unconditional measurement process, any measurement time dependence in $\hat{\mathcal{C}}_{t_{\rm m}}$ will be due to changes in the back action of JPM measurement on the cavity that change the post-measurement cavity state, and therefore modify the coherence of the post-measurement qubit state. However, for $\al_1 \gg \al_0$ sufficient decoherence of the qubit state has occurred during the drive stage that the resultant measurement time dependence of the fidelity is several orders of magnitude smaller than the average value, and in practice we can set $\mathcal{F}_{t_{\rm m}} = \mathcal{F}_{\infty}$. Using Eqs. (\ref{eqn:psi0}) and (\ref{eqn:psi1}) we can calculate $\mathcal{F}_{\infty}$ analytically in the ideal case when $\yy_{\rm R}, \yy_{\rm D} = 0$ (see Appendix \ref{app:Fid}):
\begin{align}
&\mathcal{F}_{\infty} = \frac{1}{2}\left(1+\sqrt{1-K(\al_0,\al_1)^2}\right) \label{eqn:Fidan} \\
&\nonumber K(\al_0,\al_1) = e^{-\frac{1}{2}\left(\abs{\al_0}^2 + \abs{\al_1}^2\right)} \\
&\nonumber+ \sqrt{\left(1-e^{-\abs{\al_0}^2}\right)\left(1-e^{-\abs{\al_1}^2}\right)}.
\end{align}
For $\abs{\al_0}^2 = 0$ and $\abs{\al_1}^2 = 4$, we already have $\mathcal{F}_{\infty} > 99\%$. For $\abs{\al_1}^2 = 10$ as used elsewhere, $\mathcal{F}_{\infty} > 99.99\%$. Ultimately, this value of the fidelity should be considered the fundamental limit of our protocol as it corresponds to the ideal case, ignoring both JPM relaxation and dark counts, as well as other environmental interactions.

When JPM relaxation is non-negligible ($\yy_{\rm R} \neq 0$), even for $t_{\rm m} \rightarrow \infty$ the measurement conditionally projects the qubit onto mixed states rather than the pure states of Eqs. (\ref{eqn:psi0}) and (\ref{eqn:psi1}), as even for $\al_0 = 0$, $\ket{\psi_0}$ is mixed incoherently with a $\ket{1}\bra{1}_{\rm Q}$ component. Similarly, dark counts cause mixing of the state $\ket{\psi_1}$ as they incoherently add a $\ket{0}\bra{0}$ component to $\ket{\psi_1}$. Therefore, to describe back action, we can use POVM (positive-operator valued measure) elements for the qubit state to describe the map onto the post-measurement state. While full determination of these POVM elements is beyond the scope of this work, the unconditional quantum process $\hat{\mathcal{C}}_{t_{\rm m}}$ is also directly affected by changes in the POVM elements, such that it is quantitatively different when $\yy_{\rm R}, \yy_{\rm D} \neq 0$. However, the average value of the fidelity is nearly the same, and as changes to the fidelity with measurement time are several orders of magnitude smaller than the average value, the fidelity is not a good measure to compare the ideal case with that for $\yy_{\rm R}, \yy_{\rm D} \neq 0$.

Therefore, to qualitatively study the deviations from QNDness introduced by JPM relaxation and dark counts, we examine the probability that repeated measurements (within qubit $T_1$) will give the same measurement result. Consider single measurement probabilities $P_a$, where $a\in\{0,1\}$ is the measurement outcome, and joint measurement probabilities $P_{ab}$ where $a,b$ are the outcomes of the second and first measurements, respectively. For an ideal QND measurement as defined above, we have
\begin{align}
\nonumber&P_{00} = P_{0} \\
\nonumber&P_{01} = P_{10} = 0 \\
&P_{11} = P_{1}.
\end{align}
When JPM relaxation and dark counts are taken into account, none of these relationships hold. This is generally a result of the fact that our protocol can misidentify the qubit state (due to dark counts, energy relaxation, or less than unit contrast), so that the second event does not occur with unit probability. In particular, due to dark counts $P_0 \neq P_{00}$; similarly, due to the probability of not detecting a photon for a given cavity state or mistakenly measuring the vacuum component of the $|\alpha_1\rangle$ state, we have $P_{11} \neq P_1$. $P_{11}$ can also be further modified by imperfect reset of the cavity. The symmetry between $P_{01}$ and $P_{10}$ is not broken by misidentification; however, they are both nonzero. On top of these misidentifications, QNDness can also be limited by corrections beyond the dispersive Hamiltonian as discussed in the next section.

\section*{{\normalsize{\textbf{V. Environmental Interactions and Corrections Beyond the Dispersive Hamiltonian}}}}

So far the discussion has focused on a closed qubit-cavity subsystem. When we consider interactions with the environment, it is apparent that the dominant effects are qubit and cavity relaxation. The timescale of these effects depends heavily on the frequency of the JPM, as it is most strongly coupled to environmental modes.

During the drive and reset stages of the measurement protocol, the JPM idles at a frequency that is far blue detuned from the cavity or qubit resonances. As a result, the leading order decay channel for both the cavity and the qubit is through the cavity's input port. If we take a cavity decay rate $\kappa \approx 100$ kHz, we find a qubit lifetime limitation of $T_{1}^{\kappa} \approx 2$ ms for vacuum in the cavity \cite{Beaudoin11} (see Appendix \ref{app:Decay} for further details), and this $T_{1}^{\kappa}$ will in fact increase with higher cavity occupation \cite{Sete:2014fk}, though this is a higher order effect not considered in our evaluation. This Purcell limited qubit $T_1$ can only be calculated by first considering the full Jaynes-Cummings Hamiltonian when deriving the master equation for the coupled system, and so it is inherently not contained in the dispersive picture. Essentially, induced qubit $T_1$ can be understood by observing that the eigenstates of the full JC Hamiltonian are always dressed (albeit weakly at strong detuning), and thus decay of the dressing cloud can lead to decay of the eigenstate.

In addition, while spontaneous emission of the cavity through its input port is inconsequential, emission toward the JPM during the drive stage will degrade the preparation of the cavity pointer states, with the dominant effect being a nonzero occupation $|\al_0|^2$ of the $\ket{0}_{\rm Q}$-state pointer upon completion of the drive stage. However, this effect is very small due to the large cavity-JPM detuning during the drive stage, and so it only minimally affects the contrast.

During the measurement stage, the JPM is brought on resonance with the cavity, and cavity decay through the JPM is desirable, since it amounts to bright tunnelling or JPM relaxation. However, as the cavity-JPM states hybridize, qubit decay through the JPM is also possible, as a result of beyond-dispersive effects identical to those for qubit Purcell decay discussed previously. Through a procedure similar to that of \cite{Beaudoin11}, we obtain a JPM-limited qubit lifetime of $T_1^{\gamma_{\rm R}} \approx 2\  \mu$s during the measurement stage for vacuum in the cavity, considerably shorter than $T_{1}^{\kappa} \approx 2$ ms (see Appendix \ref{app:Decay} for further details). For an occupied cavity the situation is more complex, due to additional excitations as well as stimulated emission channels, but we find that to lowest order in $g_{\rm Q}/\Delta$ the qubit lifetime increases as cavity occupation increases, and that for $\abs{\al}^2 = 10$ one would expect a qubit lifetime of $T_1^{\gamma_{\rm R}} \approx 40\  \mu$s.

\begin{figure}
\subfigure{
\label{fig:JCQubit}
\includegraphics[width = \columnwidth]{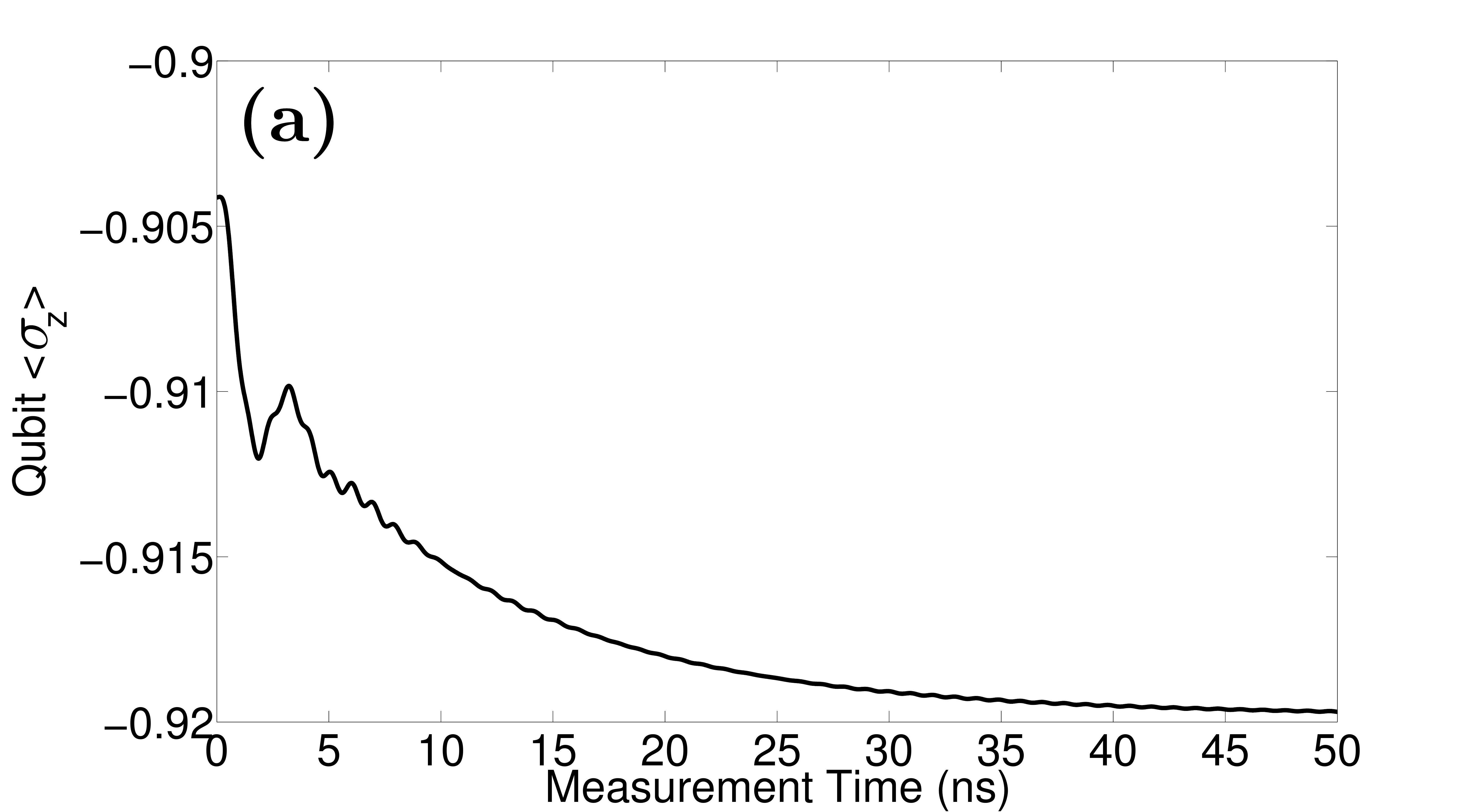}}
\subfigure{
\label{fig:Fid}
\includegraphics[width = \columnwidth]{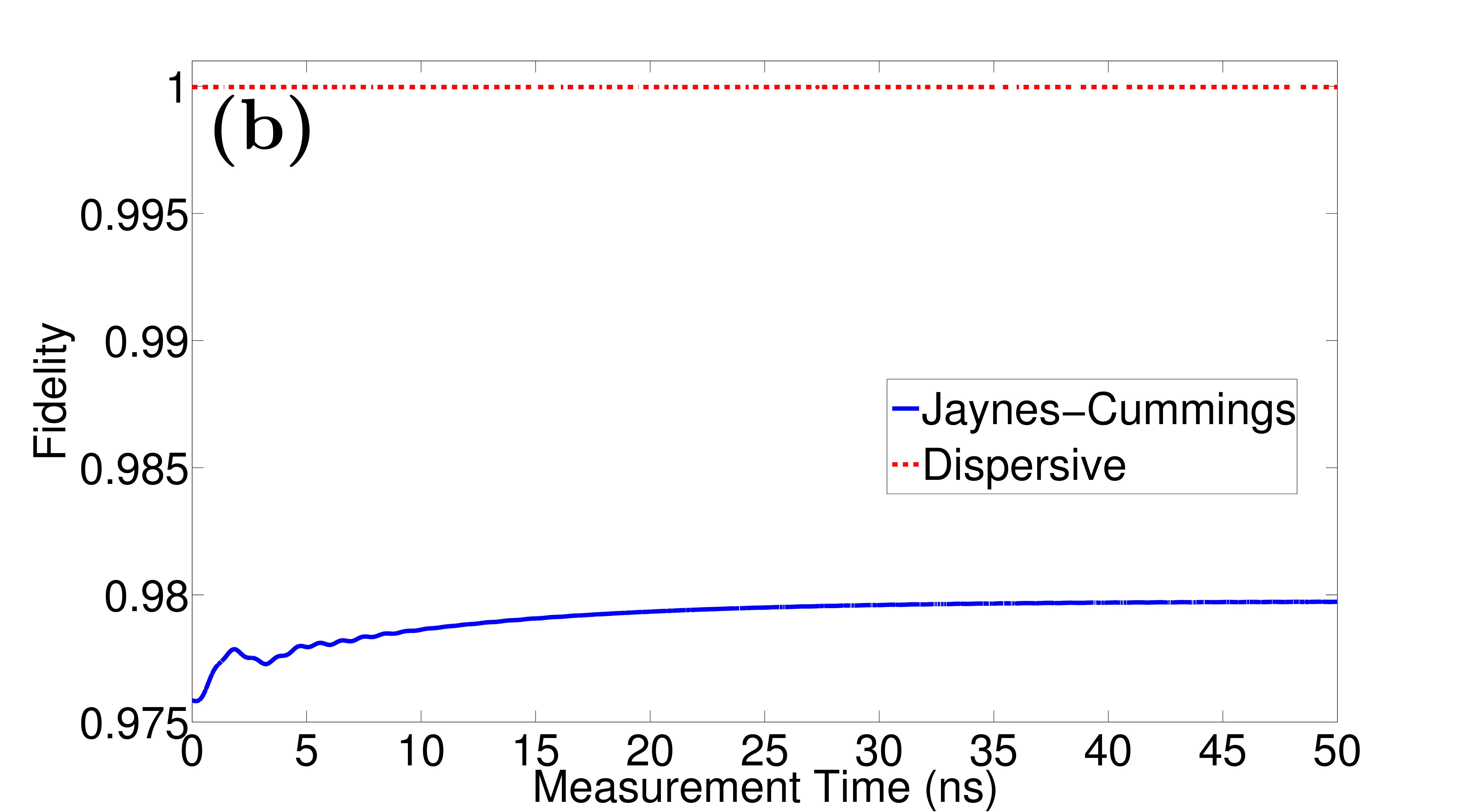}}
\caption{(Color online) {\bf (a)} Qubit $\sigma_z$ expectation for the qubit initially in the excited state versus measurement time (Jaynes-Cummings Hamiltonian) and {\bf (b)} process fidelity of qubit readout for both the dispersive Hamiltonian and the Jaynes-Cummings Hamiltonian. Coupling strength and tunnelling rates are as used throughout. Here $\abs{\al_0}^2 =0$ and $\abs{\al_1}^2 =9$.}
\end{figure}

However, numerical simulations (see Fig. \ref{fig:JCQubit}) indicate that there is no appreciable qubit decay probability during the overall measurement process. We attribute this to the fact that the global state of the system is frozen once the JPM is in the measured state, and since $\yy_{\rm J} \gg 1/T_{1}^{\yy_{\rm R}}$, this occurs long before any appreciable qubit decay. In fact, we expect only a $0.05\%$ change in the qubit state due to JPM-mediated decay (see Appendix \ref{app:Decay} for further details), which is completely washed out by other effects in Fig. \ref{fig:JCQubit}. In other words, a seemingly short induced lifetime during the measurement stage is inconsequential if the associated relaxation channel is only open for a short time. This implies that a working point with a very fast bright tunnelling rate is optimal.

The qubit also experiences dephasing due to low frequency noise at the JPM with a characteristic timescale $T_{\phi}^{\gamma_{\rm J}}$. However, as the ideal cQED measurement protocol should maximally dephase the qubit state, this low frequency noise does not affect the fidelity or measurement contrast of our protocol.

To quantify measurement degradation due to beyond-dispersive effects, we compare the process fidelity of Eq. (\ref{eqn:fid}) for dispersive qubit-cavity coupling with that for the full Jaynes-Cummings Hamiltonian. Figure \ref{fig:Fid} shows the fidelity as a function of measurement time, for similar parameters as used throughout and $\abs{\al_1}^2 = 9$. As expected, the dispersive fidelity changes only minimally as a function of measurement time, while the Jaynes-Cummings fidelity both oscillates and grows with measurement time. Crucially, the fidelity for the full Jaynes-Cummings Hamiltonian is still approximately 98\%, i.e., not significantly less than for the dispersive Hamiltonian. It is the focus of future study to improve this number.

\begin{figure}
\subfigure{
\label{fig:JCdrive}
\includegraphics[width = \columnwidth]{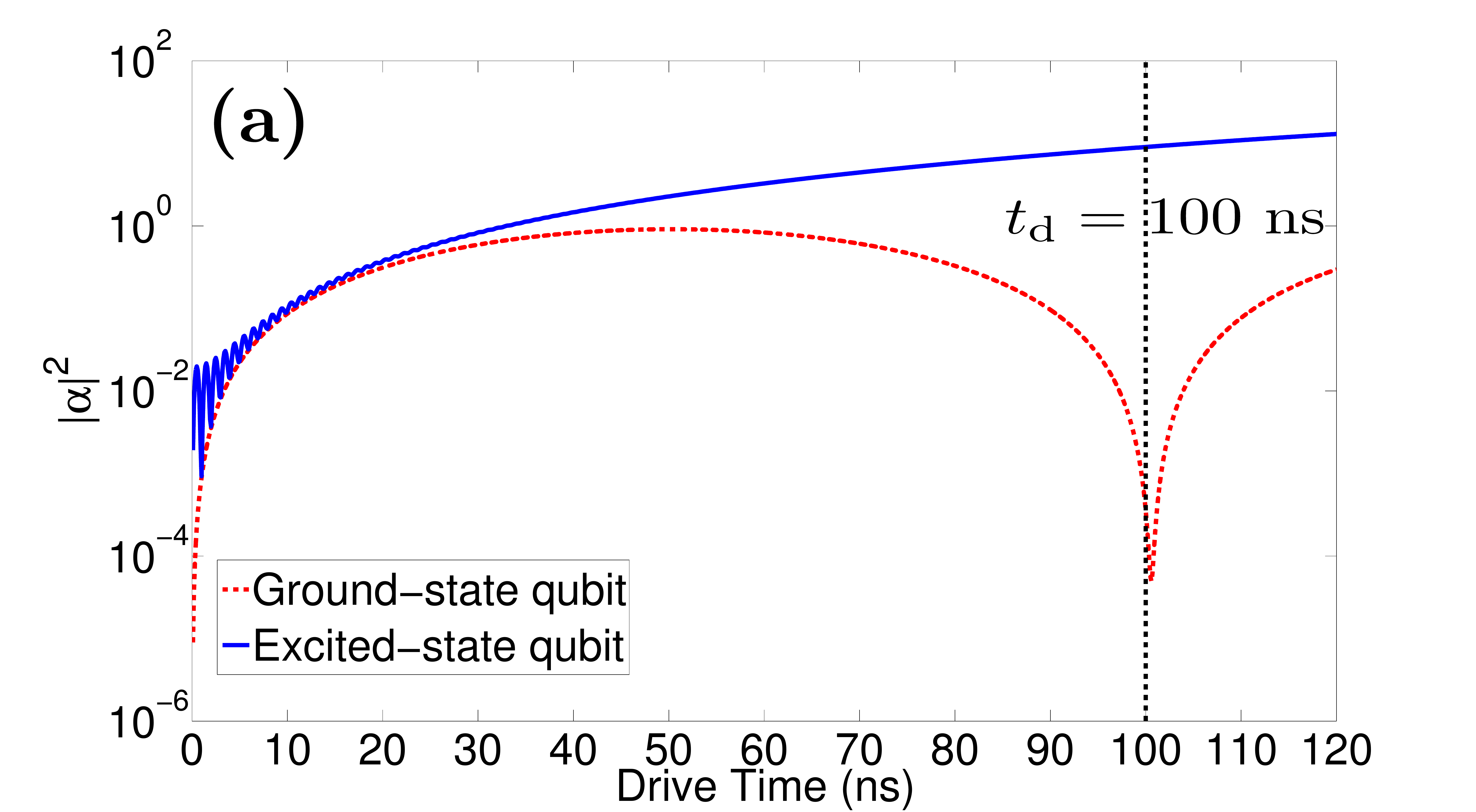}}
\subfigure{
\label{fig:JCcont}
\includegraphics[width = \columnwidth]{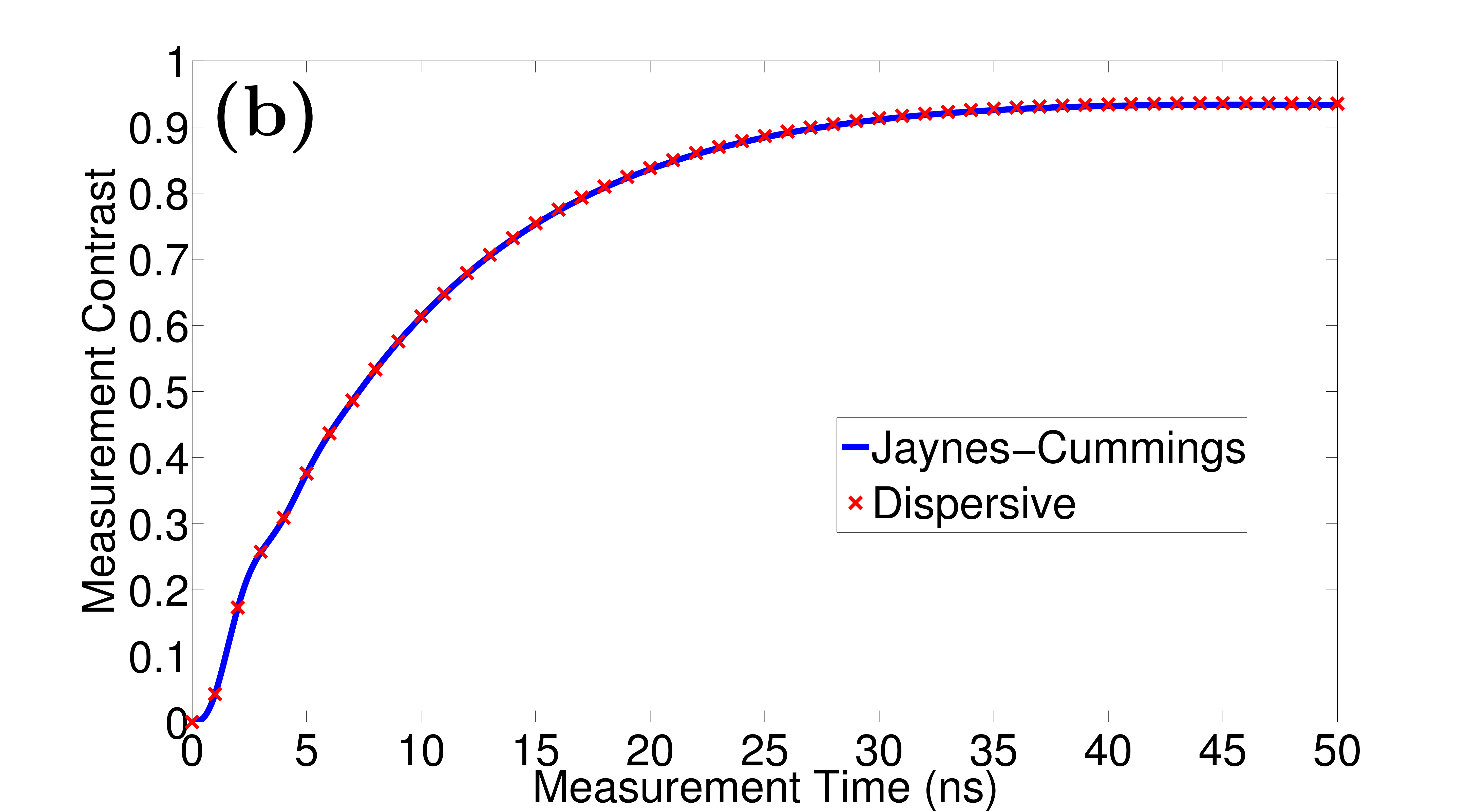}}
\caption{(Color online) {\bf (a)} Cavity occupation during the drive stage for the full Jaynes-Cummings Hamiltonian. {\bf (b)} Measurement contrast for the dispersive Hamiltonian and the Jaynes-Cummings Hamiltonian. As elsewhere, $\chi_{\rm Q}/\pi = 10$ MHz, $g_{\rm J}/2\pi  = 50$ MHz, $\yy_{\rm J}  = 200$ MHz, $\yy_{\rm D}  = 1$ MHz and $\yy_{\rm R}  = 200$ MHz. In both plots, the drive strength is chosen such that $\abs{\al_1}^2 = 9$ for $t_{\rm d} = 100$ ns for the dispersive Hamiltonian.}
\end{figure}

Finally, we have examined both the cavity occupation during the drive stage and the measurement contrast for the full Jaynes-Cummings Hamiltonian. For the drive stage the major effect is a small shift in the time $t_{\rm d}$ at which the cavity occupation is minimized for the qubit in the ground state, and an increase in the minimum occupation $\abs{\al_0}^2$ (Fig. \ref{fig:JCdrive}). This results in a reduction of the contrast (as can be seen Fig. \ref{fig:JCcont}); however, this reduction in contrast is not significant enough to seriously degrade the success of our measurement protocol.

\section*{{\normalsize{\textbf{VI. Scalability of Counting Measurement}}}}

A useful multiqubit processor comprising hundreds if not thousands of qubits will require a large number of measurement channels with their associated wiring, filtering, and isolation. It is therefore important not only to examine the ultimate performance of a single measurement channel, but also to consider prospects for scaling to many measurement channels. From the standpoint of scalability, JPM-based counter measurement possesses several unique advantages compared to conventional heterodyne measurement based on low-noise superconducting amplifiers.

The JPM requires only relatively low-bandwidth dc wiring for biasing, thus eliminating the need for bulkier cryogenic coaxial lines and microwave attenuators. Moreover, operation of the JPM requires no microwave pump tone, eliminating a major source of cost, complexity, and deleterious crosstalk in conventional cQED circuits. In addition, because the output signal of the JPM is of the order of twice the superconducting gap, no cryogenic amplifiers are needed and the JPM signal can be detected with straightforward room-temperature electronics. Alternatively, the binary digital output of the JPM provides a natural interface to the SFQ-logic family. Here, classical bits are stored in the form of quantized voltage pulses whose time integral equals the superconducting flux quantum $\Phi_0 = h/2e$. Optimized SFQ circuits can be clocked at 100s of GHz, and they offer orders of magnitude lower dissipation than conventional CMOS logic. The integration of a classical SFQ control circuit in the multiqubit cryostat would yield significant reductions in power consumption, latency, and overall system footprint.

The large intrinsic bandwidth of the JPM (approaching 1 GHz) \cite{Poudel:2012uq} also allows for the possibility of time-domain multiplexing. Multiple qubits, each with a separate readout cavity at slightly different frequencies, could be interrogated with a single JPM by selectively addressing each cavity with drive pulses at different frequencies. While the ``click''/``no click'' output of the JPM does not enable frequency-domain multiplexing of the cavity readout, it is possible to multiplex instead by staggering the readout of the cavities in time, with an offset between cavity measurements of order 10s of ns.

\section*{{\normalsize{\textbf{VII. Conclusion}}}}

In conclusion, we have outlined a new readout scheme for superconducting quantum bits using selective cavity ring-up and photodetection. We show that even without detailed optimization, our measurement protocol is compatible with the requirements of fault tolerance, with achievable measurement contrast greater than 95\% in measurement times of order 100 ns. Counter-based qubit measurement possesses distinct advantages in terms of scalability, with simple wireup and dc biasing requirements and the prospect of multiplexing in the time domain. Finally, as the counter maps quantum information to a binary digital output without the need for room-temperature heterodyne detection and post-processing, our scheme provides a natural interface between a superconducting quantum processor and cryogenic classical control circuitry based on the SFQ digital logic family.

\section*{{\normalsize{\textbf{Acknowledgments}}}}

Supported by the Army Research Office under contract W911NF-14-1-0080. LCGG, EJP, and FKW also acknowledge support from the European Union through ScaleQIT and LCGG from NSERC through an NSERC PGS-D. CX, MV and RM were also supported by NSF grant No. DMR-1105178.

\appendix
\section{Derivation of the Analytic Expression for the Bright Count Rate}
\label{app:BR}

Following [\onlinecite{Poudel:2012uq}], we begin by assuming we have a dark count-free JPM coupled to a cavity in an $N$-photon Fock state. A single photon in the cavity would cause a bright count with probability $P_1=\yy_{\rm J}/(\yy_{\rm R}+\yy_{\rm J})$, where $\yy_{\rm J}$ is the bright tunnelling rate and $\yy_{\rm R}$ is the inelastic relaxation rate of the JPM, as defined before. However, if instead the JPM relaxes back to the ground state, the second photon in the cavity will cause a bright count with probability $P_2=[\gamma_{\rm R}/(\yy_{\rm R}+\yy_{\rm J})]P_1$, where the first factor is the probability that the first photon is lost due to inelastic relaxation. Therefore, for the $n$th photon, we have
$$
P_n=\left(\frac{\yy_{\rm R}}{\yy_{\rm R}+\yy_{\rm J}}\right)^{n-1}\frac{\yy_{\rm J}}{\yy_{\rm R}+\yy_{\rm J}}.
$$
Summing up all probabilities for $n=1,\dots,N$, we obtain
$$
P_N=1-\left(\frac{\yy_{\rm R}}{\yy_{\rm R}+\yy_{\rm J}}\right)^{N}
=1-\exp\left(-N\ln\left(1+\frac{\yy_{\rm J}}{\yy_{\rm R}}\right)\right).
$$
For a coherent state, we improve the estimate by averaging over $N$ for a state with given $|\alpha|^2$:
\begin{align}
\nonumber P(\abs{\alpha}^2)&=\sum \frac{|\alpha|^{2N}}{N!}e^{-|\alpha|^2}P_N\\
&=
1-\exp\left(
-|\alpha|^2 \frac{\yy_{\rm J}}{(\yy_{\rm J}+\yy_{\rm R})}
\right).
\end{align}
This analytic expression is valid for $\yy_{\rm R}\gtrsim\yy_{\rm J}$, assuming that all rates are independent of the number of photons in the cavity. This assumption implies that the photon excites the JPM faster than both rates $\yy_{\rm R}$ and $\yy_{\rm J}$, and is valid for long measurement times.

\section{Analytic Derivation of Fidelity in the Ideal Case}
\label{app:Fid}

To calculate $\mathcal{F}_{\infty}$, we begin by defining the unconditional map on the qubit state (with $\yy_{\rm D},\yy_{\rm R} = 0$) using Eqs. (\ref{eqn:psi0}) and (\ref{eqn:psi1}) by
\be
\label{eqn:Umap}
\mathcal{E}_{\infty}\left(\ket{\psi}\bra{\psi}\right) = P_0\ket{\psi_0}\bra{\psi_0} + P_1\ket{\psi_1}\bra{\psi_1},
\ee
for an arbitrary initial qubit state $\ket{\psi} = a\ket{0} + b\ket{1}$. The ``click''/``no click'' probabilities for such an input state are given by
\begin{align}
&P_0 = \abs{a}^2e^{-\abs{\al_0}^2} + \abs{b}^2e^{-\abs{\al_1}^2}, \\
&P_1 = \abs{a}^2(1-e^{-\abs{\al_0}^2}) + \abs{b}^2(1-e^{-\abs{\al_1}^2}).
\end{align}
In light of this, Eq. (\ref{eqn:Umap}) becomes
\begin{align}
&\mathcal{E}_{\infty}\left(\ket{\psi}\bra{\psi}\right) = \abs{a}^2\ket{0}\bra{0} + \abs{b}^2\ket{1}\bra{1}\\
&\nonumber+ ab^{*}K(\al_0,\al_1)\ket{0}\bra{1} + a^{*}bK(\al_0,\al_1)\ket{1}\bra{0},
\end{align}
where $K(\al_0,\al_1)$ is as defined in Eq. (\ref{eqn:Fidan}):
\be
\nonumber K(\al_0,\al_1) = e^{-\frac{1}{2}\left(\abs{\al_0}^2 + \abs{\al_1}^2\right)}+ \sqrt{\left(1-e^{-\abs{\al_0}^2}\right)\left(1-e^{-\abs{\al_1}^2}\right)}.
\ee

Now that the map is fully determined, we can calculate the Choi matrix elements
\be
\left[\hat{\mathcal{C}}_{\infty}\right]_{ijkl} = \bra{j}\mathcal{E}_{\infty}\left(\ket{i}\bra{k}\right)\ket{l},
\ee
and find that the Choi matrix is given by
\begin{align}
\hat{\mathcal{C}}_{\infty}=\left(\begin{array}{cccc}
1 & 0 & 0 & K(\al_0,\al_1) \\
0 & 0 & 0 & 0 \\
0 & 0 & 0 & 0 \\
K(\al_0,\al_1) & 0 & 0 & 1
\end{array}\right).
\end{align}
By a similar procedure, the Choi matrix for perfect QND measurement is given by
\begin{align}
\hat{\mathcal{C}}_{\rm per}=\left(\begin{array}{cccc}
1 & 0 & 0 & 0 \\
0 & 0 & 0 & 0 \\
0 & 0 & 0 & 0 \\
0 & 0 & 0 & 1
\end{array}\right).
\end{align}
With both Choi matrices defined, using Eq. (\ref{eqn:fid}) we can calculate the fidelity to be
\be
\mathcal{F}_{\infty} = \frac{1}{2}\left(1+\sqrt{1-K(\al_0,\al_1)^2}\right),
\ee
as in Eq. (\ref{eqn:Fidan}).

\section{Cavity and JPM-Limited Qubit Lifetimes}
\label{app:Decay}

We first summarize the results of \cite{Beaudoin11} for a qubit coupled dispersively to a cavity with decay rate $\kappa$; next we extend this result to include the JPM. For a qubit coupled dispersively to a cavity, the dressed qubit-cavity eigenstates to second order are
\begin{align}
&\overline{\ket{1,n-1}} \approx \left(1 - \frac{g_{\rm Q}^2n}{2\Delta^2}\right)\ket{1,n-1} - \frac{g_{\rm Q}\sqrt{n}}{\Delta}\ket{0,n}, \\
&\overline{\ket{0,n}} \approx \left(1 - \frac{g_{\rm Q}^2n}{2\Delta^2}\right)\ket{0,n} + \frac{g_{\rm Q}\sqrt{n}}{\Delta}\ket{1,n-1},
\end{align}
where $\ket{0/1,n}$ are the uncoupled eigenstates of the cavity-qubit system and $\Delta = \ww_{\rm C} - \ww_{\rm Q}$. We are interested in Purcell-limited qubit relaxation, i.e. transitions from $\overline{\ket{1,n}}$ to $\overline{\ket{0,n}}$ mediated by the cavity's coupling to the external environment, which we assume takes the standard form with the cavity coupling operator given by $\hat{X} = \hat{a} + \hat{a}^\dagger$. From \cite{Beaudoin11}, the decay rate for this process is given by
\be
\Gamma_{\kappa}^{\overline{1n},\overline{0n}} = \kappa(\Delta_{\overline{0n},\overline{1n}})\abs{\overline{\bra{1,n}}\hat{X}\overline{\ket{0,n}}}^2;
\ee
here $\kappa(\omega)$ is the coupling constant that depends on the spectral density of the cavity's environment, which should be evaluated at $\Delta_{\overline{n0},\overline{n1}} = \ww_{\rm Q} + \chi_{\rm Q}$. To lowest order in $g_{\rm Q}/\Delta$,
\be
\abs{\overline{\bra{1,n}}\hat{X}\overline{\ket{0,n}}}^2 \approx \frac{g_{\rm Q}^2}{\Delta^2}.
\ee
Assuming an Ohmic spectral density and using as a reference value the coupling constant at the uncoupled cavity frequency $\kappa(\ww_{\rm C})$, we have $\kappa(\ww) \approx \kappa(\ww_{\rm C})\ww/\ww_{\rm C}$, where $\ww_{\rm C}$ is the bare cavity frequency, not the ultraviolet cutoff frequency of the Ohmic spectral density \cite{Weiss2012}. Setting $\kappa(\ww_{\rm C}) = 100$ kHz (the input coupling to the cavity), $\Delta/2\pi = 1$ GHz, and a corresponding $g_{\rm Q}$ that gives $\chi_{\rm Q}/2\pi = 5$ MHz, we obtain a Purcell limited qubit lifetime of 2 ms. To next order in $g_{\rm Q}/\Delta$ the qubit lifetime is dependent on the cavity occupation; however, as shown in \cite{Sete:2014fk}, the lifetime increases for higher photon numbers in the cavity.

When the JPM is brought on resonance with the cavity, the cavity-JPM states hybridize, so that the eigenstates are now
\begin{align}
\ket{n,a} = \frac{1}{\sqrt{2}}\left(\ket{n,0} + (-1)^a\ket{n-1,1} \right),
\end{align}
where the ground state $\ket{0,a} = \ket{0,0}$, and the index $a$ labels what is normally labelled by $\pm$. In light of this, to study qubit relaxation via the JPM, we must examine transitions from the two states $\overline{\ket{1,n,a}}$ to the two states $\overline{\ket{0,n,b}}$ via the JPM-environment coupling operator $\hat{\sigma}_{x}^{\rm J}$, where now
\begin{align}
&\overline{\ket{1,n,a}} \\
&\nonumber \approx \left(1 - \frac{g_{\rm Q}^2(n+1)}{2\Delta^2}\right)\ket{1,n,a} - \frac{g_{\rm Q}\sqrt{n+1}}{\Delta}\ket{0,n+1,a}, \\
&\overline{\ket{0,n,b}} \\
& \nonumber \approx \left(1 - \frac{g_{\rm Q}^2n}{2\Delta^2}\right)\ket{0,n,b} + \frac{g_{\rm Q}\sqrt{n}}{\Delta}\ket{1,n-1,b}.
\end{align}

Similar to the case for cavity-mediated decay, the decay rates for these processes are given by
\be
\Gamma_{\yy_{\rm R}}^{\overline{1na},\overline{0nb}} = \yy_{\rm R}(\Delta_{\overline{0na},\overline{1nb}})\abs{\overline{\bra{1,n,a}}\hat{\sigma}_{x}^{\rm J}\overline{\ket{0,n,b}}}^2,
\ee
where $ \yy_{\rm R}(\ww)$ is the JPM's coupling constant with the environment. Using the fact that
\begin{align}
&\bra{n+1,a}\sigma_{x}^{\rm J}\ket{n,b} = \frac{(-1)^a}{2}, \ \ &n > 0 \\
&\bra{1,a}\sigma_{x}^{\rm J}\ket{0,b} = \frac{(-1)^a}{\sqrt{2}}, \ \ &n = 0
\end{align}
and all other matrix elements are zero, we find that
\begin{align}
&\overline{\bra{1,n,a}}\hat{\sigma}_{x}^{\rm J}\overline{\ket{0,n,b}} = (-1)^a\frac{g_{\rm Q}}{2\Delta}\left(\sqrt{n}-\sqrt{n+1}\right) \\
&\overline{\bra{1,0,a}}\hat{\sigma}_{x}^{\rm J}\overline{\ket{0,0,b}} = (-1)^a\frac{g_{\rm Q}}{\sqrt{2}\Delta},
\end{align}
to first order in $g_{\rm Q}/\Delta$. As before, assuming an Ohmic spectral density we approximate $\yy_{\rm R}(\Delta_{\overline{0na},\overline{1nb}})$ by $\yy_{\rm R}(\ww_{\rm J})\ww_{\rm Q}/\ww_{\rm J}$,  which assumes that the qubit energy-shifts due to both the cavity and the JPM are sufficiently smaller than $\ww_{\rm Q}$ (i.e. $\Delta_{\overline{0na},\overline{1nb}} \approx \ww_{\rm Q}$). Using $\yy_{\rm R}(\ww_{\rm J}) = 200$ MHz as in the main text and other quantities as before, we obtain a JPM-limited qubit lifetime $T_{1}^{\yy_{\rm R}} \approx 2\  \mu$s for vacuum in the cavity. For cavity Fock states with $n>0$ this lifetime scales as
\be
T_{1}^{\yy_{\rm R}} \propto \frac{1}{\left(\sqrt{n}-\sqrt{n+1}\right)^2} = \left(\sqrt{n}+\sqrt{n+1}\right)^2, 
\ee
and so we have $T_{1}^{\yy_{\rm R}}\propto 2n$ to leading order. Thus, for $n =10$ we have an improvement of the qubit lifetime to $T_{1}^{\yy_{\rm R}} \approx 40\  \mu$s.

In addition, there are several competing incoherent processes in the JPM, namely bright and dark counts, which block the JPM-mediated qubit decay channel. Numerical simulations indicate that due to the fact that $\yy_{\rm J} \gg \Gamma_{\yy_{\rm R}}^{\overline{1na},\overline{0nb}}$ for relevant photon numbers in the cavity, there is almost no appreciable decay of the qubit during the measurement process (see Fig. \ref{fig:JCQubit}). This can be understood by the fact that $\Gamma_{\yy_{\rm R}}^{\overline{1na},\overline{0nb}}/\yy_{\rm J} \propto g_{\rm Q}^2/(n\Delta^2)$, which for $n = 10$ is only $0.05\%$, and so we expect no more than a $0.05\%$ change in the qubit state due to JPM-mediated decay during the measurement protocol. Detailed study of qubit decay during JPM measurement will be the subject of future work.

\bibliography{Counter.bib}

\end{document}